\documentclass[12pt]{article}%
\usepackage{graphicx}
\usepackage[english]{babel}
\usepackage{amsmath,amssymb,amsthm}

\begin{document}

\begin{center}{\large\bf Finite Mathematics, Finite Quantum Theory And A Conjecture On The Nature Of Time} \end{center}

\vskip 1em \begin{center} {\large Felix M. Lev} \end{center}
\vskip 1em \begin{center} {\it Artwork Conversion Software Inc.,
1201 Morningside Drive, Manhattan Beach, CA 90266, USA
(Email:  felixlev314@gmail.com)} \end{center}

\begin{abstract}
We first give a rigorous mathematical proof that classical mathematics (involving such notions as infinitely small/large, continuity etc.) is a special degenerate case of finite one in the formal limit when the characteristic $p$ of the field or ring in finite mathematics goes
to infinity.  We consider a finite quantum theory (FQT) based on finite mathematics and prove that standard continuous  quantum theory is a special case of FQT in the formal limit $p\to\infty$.  The description of states in standard quantum theory
contains a big redundancy of elements: the theory is based on real numbers while  with any desired accuracy the states can be
described by using only integers, i.e. rational and real numbers play only auxiliary role. Therefore, in
FQT infinities cannot exist in principle, FQT is based on a more fundamental mathematics than standard quantum theory
and the description of states in FQT is much more thrifty than in standard quantum theory. Space and time are purely classical notions and are not present in FQT at all. In the present paper we discuss how classical equations of motions arise as a consequence of the fact that $p$ changes, i.e. $p$ is the evolution parameter. It is shown that there exist scenarios when classical equations of motion for cosmological acceleration and gravity can be formulated exclusively in terms of quantum quantities without using space, time and standard semiclassical approximation.

\end{abstract}

\begin{flushleft} Keywords: finite mathematics, quantum theory, equations of motion\end{flushleft}

\section{Introduction}

Finite mathematics is a branch of mathematics which contains theories involving only finite sets. 
In particular, those theories cannot involve even the set of all natural numbers, to say nothing about the set of
all rational numbers and the set of all real numbers (because those sets are infinite). Known examples are theories of finite fields and finite rings (see e.g. textbooks in Ref. \cite{textbooks}).  In Sec. \ref{mainstatement} we give a rigorous mathematical proof that finite mathematics is fundamental and classical mathematics (involving such
notions as infinitely small/large, continuity etc.) is a special degenerate  case of finite one in the formal limit
when the characteristic $p$ of the field or ring in finite mathematics goes to infinity. 
This fact is very important not only for constructing
fundamental quantum theory but also for mathematics itself. We consider
a version of quantum theory called finite quantum theory (FQT) where
quantum states are elements of a linear space over a finite field or ring with a characteristic $p$,
and operators of physical quantities are linear operators in this space.  In Sec. \ref{mainstatement} we also prove
that standard continuous quantum theory is a special degenerate case of FQT in the formal limit $p\to\infty$.
 As follows from the proof, the description of states in standard quantum theory
contains a big redundancy of elements: with any desired accuracy the states can be
described by using only integers, i.e. rational and real numbers play only auxiliary
role and are not needed for this description. Therefore, in addition to the fact that
FQT is based on more fundamental mathematics, the description of states in FQT is
also much more thrifty than in standard quantum theory. We believe that those results
can be understood even by physicists who are not familiar with basic facts of finite mathematics.

As shown in Refs. \cite{lev4,monograph,lev2}, FQT sheds a new light on fundamental problems of gravity and particle theory. In the present paper
we discuss how FQT can be applied for solving such a fundamental problem of quantum theory as the problem of
time. In Sec. \ref{intro} we argue that space and time are pure classical notions, and even standard quantum theory
should not involve those notions. However, since quantum theory is treated as more fundamental than classical theory,
quantum theory should explain how classical equations of motion arise on the basis of this theory.

In Sec. \ref{symmetry} we argue that quantum theory should proceed not from classical space-time but from
symmetry on quantum level proposed by Kondratyuk.  In Secs. \ref{2particles} - \ref{S9} we consider a system
of two free particles in quantum theory based on de Sitter symmetry. We consider both, the case of standard quantum
theory and the case when quantum theory is based on a finite ring or field with characteristic $p$. 
It is shown that the cosmological acceleration
is simply a kinematical consequence of de Sitter symmetry, and gravity can be treated as  a kinematical consequence
of a symmetry based on finite mathematics.

In Sec. \ref{classicalFQT} we discuss our conjecture that classical time manifests itself as a consequence of the fact
that the characteristic $p$ changes. It is shown that there exist scenarios when classical equations of motion for cosmological acceleration and gravity can be formulated exclusively in terms of quantum quantities without using space, time and standard semiclassical approximation.

\section{Remarks on fundamental theories}
\label{fundamentaltheories}

The notions of infinitely small/large, continuity etc.  have been proposed by Newton and Leibniz more than
300 years ago. At that times people did not know about
atoms and elementary particles. On the basis of everyday experience they believed that any macroscopic 
object can be divided into arbitrarily large number of arbitrarily small parts. However, from the point of view of 
the present knowledge those notions look problematic. 

For example, a glass of water contains approximately $10^{25}$ molecules.
We can divide this water by ten, million, etc. but when we reach the level of atoms and elementary particles
the division operation loses its usual meaning and we cannot obtain arbitrarily small parts. So, {\it any description of macroscopic phenomena using continuity and differentiability can be only approximate}. 
In nature there are no continuous curves and surfaces. For example, if we draw a line on a sheet of paper 
and look at this line by a microscope then we will see that the line is strongly discontinuous because it consists of
atoms.

Nevertheless, a belief of the overwhelming majority of scientists is that classical mathematics
is fundamental while finite mathematics 
is something inferior what is used only in special applications. In particular, the majority of physicists believe
that even ultimate quantum theory describing atoms and elementary particles will be based on classical
mathematics. 

One of the arguments is that discrete spectrum appears in many problems of 
classical mathematics and elementary particles might be
discrete solutions of some quantum field theory equations. In this scenario the notions of infinitely small, continuity
and division does not have the usual physical meaning but remain as consistent mathematical notions, although
this situation does not  seem to be natural.

In the next section we give a rigorous mathematical proof that classical mathematics is not more
fundamental than finite one but the situation is the opposite: {\it classical mathematics is a special degenerate case of finite one}. Since this point is extremely important not only for physics but even for mathematics itself, below we give a detailed motivation in favor of this statement. For this purpose we first consider three known comparisons of physical theories. 

First we compare relativistic theory (RT) with nonrelativistic one (NT). One of the reasons why RT can be treated as
more fundamental is that it contains a finite parameter $c$ and NT can be treated as a special degenerate
case of RT in the formal limit $c\to\infty$. Therefore, by choosing a large value of $c$,  RT can reproduce any result of NT with any desired accuracy. On the contrary, when the limit is already taken one cannot return back from NT to
RT and NT cannot reproduce all results of RT. It can reproduce only results obtained when $v\ll c$.  

Compare now de Sitter (dS) and anti-de Sitter (AdS) invariant theories with RT. One of the reasons why the former can be treated as
more fundamental than the latter is that they contain a finite parameter $R$ (which can be called the radius of the world)
and RT can be treated as a special degenerate
case of dS or AdS theories in the formal limit $R\to\infty$. Therefore, by choosing a large value of $R$, dS and AdS theories can reproduce any result of RT with any desired accuracy. On the contrary, when the limit is already taken one cannot return back from RT to dS and AdS theories, and RT cannot reproduce all results of those theories.   

In his famous paper "Missed Opportunities" \cite{Dyson} Dyson notes that RT is more fundamental than NT, and dS and AdS theories
are more fundamental than RT not only from physical but also from pure mathematical considerations. Poincare group is
more symmetric than Galilei one and the transition from the former to the latter at $c\to\infty$ is called contraction. Analogously dS and AdS groups are more symmetric than Poincare one and the transition from the former to the latter at $R\to\infty$ (described in Sec. \ref{symmetry}) also is called contraction. At the same time, since dS and AdS groups are semisimple they have a maximum possible symmetry and cannot be obtained from more symmetric groups by contraction.

The paper \cite{Dyson} appeared in 1972 and, in view of Dyson's remarks, a question arises why fundamental theories
of elementary particles (QED, electroweak theory and QCD) are still based on Poincare symmetry and not dS or AdS
symmetries. A typical justification is that since $R$ is much greater that dimensions of elementary particles, there is
no need to use the latter symmetries for description of elementary particles.

We believe that this argument is not consistent because usually more general theories shed a new
light on standard concepts. For example, as shown in Ref. \cite{JPA}, in contrast to the
situation in Poincare invariant theories, where a particle and its antiparticle are described by different
irreducible representations (IRs) of the Poincare algebra (or group), in dS theory a particle and its antiparticle
belong to the same IR of the dS algebra. In the formal limit $R\to\infty$ one IR of the dS algebra splits into two
{\it different} IRs of the Poincare algebra for a particle and its antiparticle. Strictly speaking, this implies that in dS
theory the very notion of a particle and its antiparticle is only approximate since transitions 
particle$\leftrightarrow$antiparticle are not prohibited. 

As a consequence, in dS theory the conservation of electric charge and
baryon and lepton quantum numbers can be only approximate. In particular, one might hypothesize that the
known phenomenon of baryon asymmetry of the Universe is a consequence of the fact that at early stages
of the Universe the value of $R$ was much less than now and for this reason the nonconservation of the
baryon quantum number was rather strong.  

At the same time, a problem arises that particles which in Poincare invariant theory
are neutral (i.e. coincide with their antiparticles) cannot be described by IRs of the dS
algebra because here the number of states in IRs is twice as big as in IRs of the Poincare algebra. This poses
a question whether neutral particles (e.g. even photons) can be elementary.

In the case of AdS symmetry the situation is similar to that in Poincare theory. However, as shown in Refs. 
\cite{lev4,monograph}, in FQT the situation is similar to that in dS theory rather than in AdS theory. 

Compare now quantum theory with classical one. One of the reasons why the former can be treated as
more fundamental is that it contains a finite parameter $\hbar$ and classical theory can be treated as a special 
degenerate case of quantum one in the formal limit $\hbar\to 0$. Therefore, by choosing a small value of $\hbar$, quantum
theory can reproduce any result of classical one with a high accuracy. On the contrary, when the limit is 
already taken one cannot return back from classical to quantum theory and the former
cannot reproduce all results of the latter.  

All the three discussed comparisons give a motivation for the following

{\bf Definition:} {\it Let theory A contain a finite parameter and theory B be obtained from theory A in the formal limit when the parameter goes to zero or infinity. Suppose that with any desired accuracy theory A can reproduce any result of theory B by choosing a value of the parameter. On the contrary, when the limit is already
taken then one cannot return back to theory A and theory B cannot reproduce all results of theory A. Then theory A is more general than theory B and theory B is a special degenerate case of theory A}. We will see below that {\bf Definition} also describes the relation between finite mathematics and classical one.

In the literature the notion of the $c\hbar G$ cube of physical theories is sometimes used. The meaning is
that any relativistic theory should contain $c$, any quantum theory should contain $\hbar$ and any gravitation 
theory should contain $G$.
The more fundamental a theory is the greater number of those parameters it contains. In particular, relativistic quantum theory of
gravity is the most fundamental because it contains all the three parameters $c$, $\hbar$ and $G$ while nonrelativistic
classical theory without gravitation is the least fundamental because it contains none of those parameters.
In our papers we argue that the $c\hbar R$ cube is more relevant than the $c\hbar G$ one, and the reasons will
be explained below.

An impression might arise that since nonrelativistic classical mechanics without gravity
does not have any of the above parameters it is the most fundamental. However, this impression is not correct because
this theory contains three parameters ($kg,m,s$). The most general dS or AdS quantum theories can be reformulated
such that they are dimensionless and do not contain the parameters  ($c,\hbar,R$) at all. Those parameters arise
only when we discuss transitions from more general theories to less general ones.

As an illustration, consider a measurement of a component of angular momentum. The
result depends on the system of units. As shown in quantum
theory, in units $\hbar/2=1$ the result is given by an integer
$0, \pm 1, \pm 2,...$. We can reverse the order of units
and say that in units where the angular momentum is an integer $l$, its
value in $kg\cdot m^2/sec$ is $(1.05457162\cdot 10^{-34}\cdot
l/2)kg\cdot  m^2/s$. Which of those two values has more
physical significance? In units where the angular momentum
components are integers, the commutation relations between the
components are
\begin{equation}
[M_x,M_y]=2iM_z,\quad [M_z,M_x]=2iM_y,\quad [M_y,M_z]=2iM_x
\label{MxMy}
\end{equation}
and do not depend on any parameters. Then the meaning of
$l$ is clear: it shows how big the angular momentum is in
comparison with the minimum nonzero value 1. At the same time,
the measurement of the angular momentum in units $kg\cdot
m^2/s$ reflects only a historic fact that at macroscopic
conditions on the Earth in the period between the 18th and 21st
centuries people measured the angular momentum in such units.

For quantum theory itself the quantity $\hbar$ is not needed. Classical theory is a good
approximation for quantum one when all angular momenta in question are very large. 
From the formal point of view $\hbar$ is needed only as a formal intermediate step for getting classical theory from quantum one:
we first write quantum theory with $\hbar$, then take the limit $\hbar\to 0$ and then in classical theory the quantity of
angular momentum has the dimension $kg\cdot m^2/s$.

Analogous remarks can be given on the
quantity $c$ (see e.g. Ref. \cite{monograph}). Nonrelativistic theory is a good approximation for
relativistic one when all velocities in question are much less than unity. Relativistic theory by itself does 
not need $c$. It is needed only as a formal  intermediate step for getting nonrelativistic theory from relativistic one: we first write relativistic theory with $c$, then take the limit $c\to\infty$ and then in nonrelativistic theory the quantity of velocity has the dimension $m/s$. The discussion of the quantities $R$ and $G$ is given below.

\section{Proof that finite mathematics is more fundamental than classical one and FQT is more
fundamental than standard quantum theory}
\label{mainstatement}

We believe that, as a preliminary step, it is important to discuss philosophical aspects of such a simple problem as operations with natural numbers.

\subsection{Remarks on arithmetic}
\label{remarks}

In the 20s of the 20th century the Viennese circle of philosophers
under the leadership of Schlick developed an approach called logical positivism which contains
verification principle:  {\it A proposition is only cognitively meaningful if it can be definitively and 
conclusively 
determined to be either true or false} (see e.g. Refs. \cite{verificationism}). 
On the other hand, as noted by Grayling
\cite{Grayling}, {\it "The general laws of science are not, even in principle, verifiable, if verifying means 
furnishing conclusive proof of their truth. They can be strongly supported by repeated experiments and 
accumulated evidence but they cannot be verified completely"}. Popper proposed the 
concept of falsificationism \cite{Popper}: {\it If no cases where a claim is false can be found, then 
the hypothesis is accepted as provisionally true}. 

According to the principles of quantum theory, there should be no statements
accepted without proof and based on belief in their correctness (i.e. axioms).
The theory should contain only those statements that can be verified, at least in principle, 
where by "verified" physicists mean 
experiments involving only a finite number of steps. So the philosophy of quantum theory is similar to 
verificationism, not falsificationism. Note that Popper was a strong opponent of the philosophy of
quantum theory and supported Einstein in his dispute with Bohr.

The verification principle does
not work in standard classical mathematics. For example, it cannot be determined whether the statement that 
$a+b=b+a$ for all natural numbers $a$ and $b$ is true or false. According
to falsificationism, this statement is provisionally true until one has found some numbers $a$ and $b$ for which $a+b\neq b+a$. There exist different theories of arithmetic
(e.g. Peano arithmetic or Robinson arithmetic) aiming to solve foundational problems of
standard arithmetic. However, those theories are incomplete and are not used in applications.

From the point of view of verificationism and principles of quantum theory, classical mathematics is not well defined
not only because it contains an infinite number of numbers. For example, let us pose a problem whether 
10+20 equals 30. Then one should describe an experiment
which gives the answer to this problem. Any computing device can operate only with a finite number of bits and can perform
calculations only modulo some number $p$. Say $p=40$, then the experiment will confirm that
10+20=30 while if $p=25$ then we will get that 10+20=5. 

{\it So the statements that 10+20=30 and even that $2\cdot 2=4$
are ambiguous because they do not contain information on how they should be verified.} 
On the other hands, the statements
$$10+20=30\,(mod\, 40),\,\, 10+20=5\,(mod\, 25),$$
$$2\cdot 2=4\,(mod\, 5),\,\, 2\cdot 2=2\,(mod\, 2)$$
are well defined because they do contain such an information.
So, from the point of view of verificationism and principles of quantum theory, only operations modulo a number are well defined.

We believe the following 
observation is very important: although classical 
mathematics (including its constructive version) is a part of our everyday life, people typically do not realize that {\it classical 
mathematics is implicitly 
based on the assumption that one can have any desired amount of resources}. 
So classical mathematics is based on the implicit assumption that we can consider an idealized case
when a computing device can operate with an infinite number of bits. In other words, standard
operations with natural numbers are implicitly treated as limits of operations modulo $p$ 
when $p\to\infty$. Usually  in mathematics, legitimacy of every limit is thoroughly investigated, 
but in the simplest
case of standard operations with natural numbers it is not even mentioned that those
operations can be treated as limits of operations modulo $p$. In real life such limits even 
 might not exist if, for example, the Universe contains a finite number of elementary particles.

Classical mathematics proceeds from standard arithmetic which does not contain operations
modulo a number while finite mathematics necessarily involves such operations. In the
next subsection we prove that, regardless of philosophical preferences, finite mathematics
is more fundamental than classical one and FQT is more fundamental than standard quantum theory. 

\subsection{Proof of the main statement}
\label{proof}

Classical mathematics starts from natural numbers but here only addition and multiplication are always possible. In order to make addition invertible we introduce negative integers and get the ring of integers $Z$. However, if instead of all natural numbers we consider only a set $R_p$ of 
$p$ numbers 0, 1, 2, ... $p-1$ where addition and multiplication are defined
as usual but modulo $p$ then we get a ring without adding new elements. In our
opinion the notation $Z/p$ for $R_p$ is not quite adequate because it may
give a wrong impression that finite mathematics starts from the infinite set $Z$
and that $Z$ is more general than $R_p$. However, although $Z$ has more elements than $R_p$, $Z$ cannot be more general than $R_p$
because $Z$ does not contain operations modulo a number.

Since operations in $R_p$ are modulo $p$, one can represent 
$R_p$ as a set $\{0,\pm 1,\pm 2,...,\pm(p-1)/2)\}$ if $p$ is odd and as a set
$\{0,\pm 1,\pm 2,...,\pm (p/2-1),p/2\}$ if $p$ is even. Let $f$ be a function 
from $R_p$ to $Z$ such that
$f(a)$ has the same notation in $Z$ as $a$ in $R_p$. 
If elements of $Z$ are depicted as integer points on the $x$ axis of the $xy$ plane then, if $p$
is odd, the elements of $R_p$
can be depicted  as points of the circumference in Fig. 1.
\begin{figure}[!ht]
\centerline{\scalebox{0.3}{\includegraphics{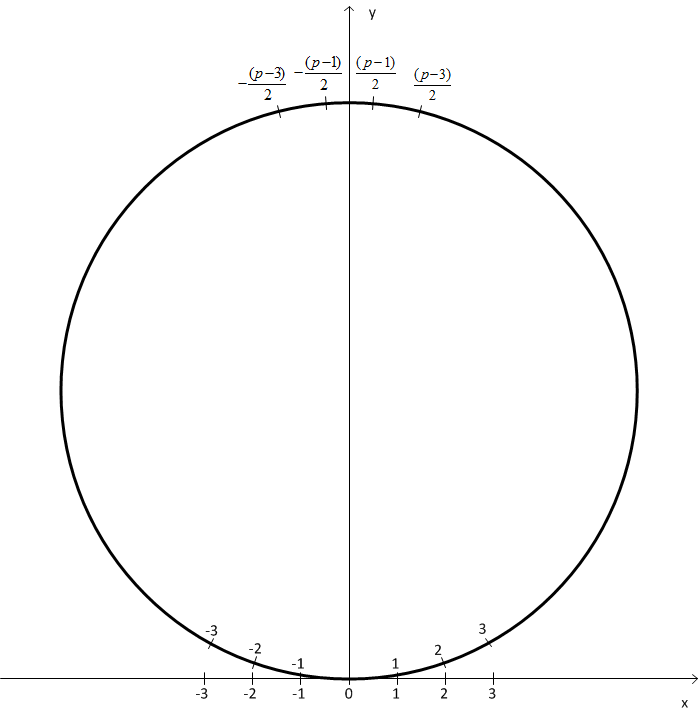}}}
\caption{
  Relation between $R_p$ and $Z$
}
\label{Fig.1}
\end{figure}
and analogously if $p$ is even. 
This picture is natural  since $R_p$ has a property that if we take any element $a\in R_p$ and sequentially add 1 then after $p$ steps we will exhaust
the whole set $R_p$ by analogy with the property that if we move along a circumference in the same direction
then sooner or later we will arrive at the initial point. 

We define the function $h(p)$ such that $h(p)=(p-1)/2$ if $p$ is odd and $h(p)=p/2-1$ if $p$ is even.
Let $n$ be a natural number and $U(n)$ be a set of elements $a\in R_p$ such that 
$|f(a)|^n \leq h(p)$. Then $\forall m \leq n$ the result of any $m$ operations of addition,
subtraction or multiplication of elements $a\in U(n)$ is the same as for the corresponding elements $f(a)$ in $Z$, i.e. in this
case operations modulo $p$ are not explicitly manifested.

Let $n=g(p)$ be a function of $p$ and $G(p)$ be a function such that the set $U(g(p))$ contains
at least the elements $\{0,\pm 1,\pm 2,..., \pm G(p)\}$. In what follows $M>0$ and $n_0>0$ are natural numbers.
If there is a sequence of natural numbers
$(a_n)$ then standard definition that $(a_n)\to \infty$ is that $\forall M$ $\exists n_0$ 
such that $a_n\geq M\,\, \forall n\geq n_0$. By analogy with this
definition we will now prove 

{\it Statement 1: There exist functions $g(p)$ and $G(p)$ such that $\forall M$ $\exists p_0>0$ such that $g(p)\geq M$ and $G(p)\geq 2^M\,\, \forall p\geq p_0$.}
 
\begin{proof}
$\forall p>0$ there exists a unique natural $n$ such that $2^{n^2}\leq h(p)<2^{(n+1)^2}$. Define
$g(p)=n$ and $G(p)=2^n$. Then $\forall M\,\, \exists p_0$ such that $h(p_0)\geq 2^{M^2}$. Then
$\forall p\geq p_0$ the conditions of {\it Statement 1} are satisfied.
\end{proof}

The problem of actual infinity is discussed in a vast literature. The {\it technique} of classical 
mathematics does not involve actual infinities and here infinities are understood only as limits. 
However, the {\it basis} of classical mathematics does involve actual infinities. 
For example, $Z$ is not treated as a limit of finite sets, i.e. it is treated as actual and not potential infinity. Moreover, classical set theory
considers infinite sets with different cardinalities.  

{\it Statement 1} is the proof
that the ring $Z$ is the limit of the ring $R_p$ when $p\to\infty$, and the result of any finite combination of additions, subtractions and multiplications in
$Z$ can be reproduced in $R_p$ if $p$ is chosen to be sufficiently large. On the contrary, when the limit is already taken then one cannot return back from $Z$ to
$R_p$, and in $Z$ it is not possible to 
reproduce all results in $R_p$ because in $Z$ there are no operations modulo a number. 
According to {\bf Definition} in Sec. \ref{fundamentaltheories} this means that {\it the ring $R_p$ is more general than $Z$, and $Z$ is a special degenerate case of $R_p$}.

When $p$ is very large then $U(g(p))$ is a relatively small part of $R_p$, and in general the results
in $Z$ and $R_p$ are the same only in $U(g(p))$. This is analogous to the fact mentioned in
Sec. \ref{fundamentaltheories} that the results of NT and RT are the same only in relatively small cases
when velocities are much less than $c$. However, 
 when the radius of the circumference in Fig. 1 becomes infinitely large then
a relatively small vicinity of zero in $R_p$ becomes the infinite set $Z$ when $p\to\infty$. 
{\it This example demonstrates that once we involve infinity and replace $R_p$ by $Z$ then we automatically 
obtain a degenerate theory because in $Z$ there are no operations modulo a number}.

In classical mathematics, the ring $Z$ is the starting point for introducing the notions of rational and real numbers. Therefore those notions arise from a degenerate set. Then a question arises whether the fact that $R_p$ is more general than Z implies
that finite mathematics is more general than classical one, i.e. whether finite 
mathematics can reproduce all results obtained by applications of classical mathematics.
For example, if $p$ is prime then $R_p$ becomes the Galois field $F_p$, and the results
in $F_p$ considerably differ from those in the set $Q$ of rational numbers even when $p$ is
very large. In particular, 1/2 in 
$F_p$ is a very large number $(p + 1)/2$.  Since quantum theory is the most
general physical theory (i.e. all other physical theories are special cases of quantum one), 
the answer to this question depends on whether standard quantum theory
based on classical mathematics is most general or is a special degenerate case of a more general 
quantum theory.

As noted in Sec. \ref{fundamentaltheories}, dS and AdS quantum theories are more general than Poincare quantum theory. In the former, quantum states are described by representations of the dS or AdS algebras, respectively. For each algebra the representation
operators of the basis elements $M^{ab}$ ($a,b=0,1,2,3,4$, $M^{ab}=-M^{ba}$) satisfy the
commutation relations
\begin{equation}
[M^{ab},M^{cd}]=-2i (\eta^{ac}M^{bd}+\eta^{bd}M^{ac}-
\eta^{ad}M^{bc}-\eta^{bc}M^{ad})
\label{CR}
\end{equation}
where $\eta^{ab}$ is the diagonal tensor such that
$\eta^{00}=-\eta^{11}=-\eta^{22}=-\eta^{33}=1$, and $\eta^{44}\pm 1$ for the dS and AdS cases,
respectively. According to principles of quantum theory, from these ten operators one should construct a maximal set $S$
of mutually commuting operators defining independent physical quantities and construct a basis in the
representation space such that the basis elements are eigenvectors of the operators from $S$.

The rotation subalgebra of algebra (\ref{CR}) is
described in every textbook on quantum mechanics. The basis of the subalgebra is $(M_x,M_y,M_z)=(M^{23},M^{31},M^{12})$ and those 
operators commute according to Eq.  (\ref{MxMy}). 
A possible choice of $S$ is $S=(M_z,K)$ where $K=M_x^2+M_y^2+M_z^2$ is the Casimir
operator of the subalgebra, i.e. it commutes with all the operators of the subalgebra. Then any IR of the subalgebra is described by an integer $k\geq 0$. The basis elements $e(\mu , k)$ of the representation space are eigenvectors of the operator $K$ with the eigenvalue $k(k+2)$ and
the eigenvectors of the operator $M_z$ with the eigenvalues $\mu$ such that, for a given $k$, $\mu$ can
take $k+1$ values $\mu=-k,-k+2,...,k-2,k$. Therefore all the basis elements are eigenvectors of the operators
from $S$ with the eigenvalues belonging to $Z$.

As shown in Ref. \cite{lev4} and in chapters 4 and 8 of Ref. \cite{monograph} 

{\it Statement 2: For algebra (\ref{CR}) there exist sets $S$ and representations such that basis vectors
in the representation spaces are eigenvectors of the
operators from $S$ with eigenvalues belonging to $Z$. Such representations reproduce
standard representations of the Poincare algebra in the formal limit $R\to\infty$}.
Therefore the remaining problem is whether or not quantum theory based
on finite mathematics can be a generalization of standard quantum theory where
states are described by elements of a separable complex Hilbert spaces $H$. 

Let $x$ be an element of $H$ and $(e_1,e_2,...)$ be
a basis of $H$ normalized such that the norm of each $e_j$ is an integer. Then {\it with any desired 
accuracy each element of $H$ can be approximated by a finite linear combination
\begin{equation}
x=\sum_{j=1}^n c_je_j 
\label{sum}
\end{equation}
where $c_j=a_j+ib_j$ and all the numbers $a_j$ and $b_j$ ($j=1,2,....n$) are
rational}. 
This follows from the known fact that the set of such sums is dense in $H$.

The next observation is that spaces in quantum theory are projective, i.e. for any complex
number $c\neq 0$ the elements $x$ and $cx$ describe the same state. This follows from the
physical fact that not the probability itself but only ratios of probabilities have a physical
meaning. In view of this property, both parts of Eq. (\ref{sum}) can be multiplied by a 
common denominator of all the numbers $a_j$ and $b_j$. As a result, we have

{\it Statement 3: Each element of $H$ can be approximated by a finite linear combination (\ref{sum})
where all the numbers $a_j$ and $b_j$ belong to $Z$}.

We conclude that Hilbert spaces in standard quantum theory contain a big redundancy of
elements. Indeed, although formally the description of states in standard quantum theory
involves rational and real numbers, such numbers play only an auxiliary role because 
with any desired accuracy each state can be described by using only integers. 
Therefore, as follows from {\bf Definition} in Sec. \ref{fundamentaltheories} and {\it Statements 1-3},
\begin{itemize}
\item Standard quantum theory based on classical mathematics is a special degenerate case
of quantum theory based on finite mathematics.
\item Even classical mathematics
itself is a special degenerate case of finite mathematics  in the formal limit when the characteristic of the field or ring in the latter goes to infinity. 
\end{itemize}
Note that the last statement is meaningful only if in applications finite mathematics is 
more pertinent  than classical one while if those theories are treated only as abstract ones
than the statement that one theory is more fundamental than the other is meaningless.

\subsection{Discussion}
\label{conclusion}

The above construction has a known historical analogy. For many years people believed
that the Earth was flat and infinite, and only after a long period of time they realized that
it was finite and curved. It is difficult to notice the curvature dealing only with
distances much less than the radius of the curvature. Analogously one might think that
the set of numbers describing nature in our Universe has a "curvature" defined by a very
 large number $p$ but we do not notice it dealing only with numbers much less than $p$. 

As noted in the preceding subsection, introducing infinity automatically implies transition to a degenerate 
theory because in this case operations modulo a number are lost. Therefore {\it even from the pure mathematical point of view} the notion of infinity cannot be fundamental, 
and theories involving infinities can be only approximations of more general theories.

In the preceding subsection we have proved that classical mathematics is a special degenerate case of finite one in the formal limit $p\to\infty$ and that FQT is more fundamental
than standard quantum theory. 
The fact that at the present stage of the Universe $p$ is a huge number explains why in many cases classical mathematics describes natural phenomena with a very high accuracy. 
At the same time, as noted below, the explanation of several phenomena can be given only
in the theory where $p$ is finite.

Although classical mathematics is a degenerate case of finite one, a problem arises
whether classical mathematics can be substantiated as an abstract science.
It is known that, in spite of great efforts of many great mathematicians, the problem 
of foundation of
classical mathematics has not been solved. For example, G\"{o}del's incompleteness theorems state that no system of axioms can ensure that all facts about natural numbers can be proven and the system of axioms in classical mathematics cannot demonstrate its own consistency. Let us recall
that classical mathematics does not involve operations modulo a number.

The philosophy of Cantor, Fraenkel, G\"{o}del, Hilbert, Kronecker, Russell, Zermelo and 
other great mathematicians was based on macroscopic experience in which the 
notions of infinitely small, infinitely large, continuity and standard division are natural. 
However, as noted above, those notions contradict the existence of elementary particles and are not natural
in quantum theory. The illusion of continuity arises when one neglects the discrete structure of matter.

However, since in applications classical mathematics is a special degenerate case of finite one,
foundational problems of classical mathematics are important only when it is treated as an abstract
science. The technique of classical mathematics is very powerful and in many cases (but not all of them) describes reality with a high accuracy.

\section{Transition from standard quantum theory to FQT}
\label{FQT}

The official birth-year of quantum theory is 1925. The meaning of "quantum" is discrete and the presence of this word in
the name of the theory reflects the fact that some quantities have a discrete spectrum. 
The founders of the theory were highly educated physicists
but they used only classical mathematics and even now mathematical education at physics departments
does not involve discrete and finite mathematics. From the formal point of view, the existence of discrete
spectrum in classical mathematics is not a contradiction. On the other hand, discrete spectrum can be treated
as more general than continuous one: the latter can be treated as a special degenerate case of the former in a special case when distances between the levels of the discrete spectrum become (infinitely) small. 

In physics there are known examples in favor of this point of view. For example, the angular momentum operator has a pure discrete spectrum which becomes the continuous one in the formal limit $\hbar\to 0$. Another example is the following. As noted
above and shown in Sec. \ref{symmetry}, Poincare symmetry is a special degenerate case of dS or AdS symmetries. The procedure when the latter becomes the former is performed as follows. Instead of some four dS or AdS angular momenta $M$ we introduce standard Poincare four-momentum $P$ such that $P= M/(2R)$ where R is a formal parameter 
which is called the radius of the world. The spectrum of the operators $M$ is discrete, the distances between the spectrum eigenvalues are of the order of $\hbar$ and therefore at this stage the Poincare four-momentum $P$ has the discrete spectrum such that the distances between the spectrum eigenvalues are of the order of $\hbar/R$. In the formal limit $R\to\infty$ the commutation relations for the dS and AdS algebras become the commutation relations for the Poincare algebra and instead of the discrete spectrum for the operators $M$ we have the continuous spectrum for the operators $P$. 
In Sec. \ref{symmetry} this transition is discussed in greater details. 

The above remarks about the discrete spectrum are valid even in standard quantum theory based on
classical mathematics. Here the state of a system is
described by a vector $\tilde x$ from a separable Hilbert space $H$. We now use a "tilde" to denote elements
of Hilbert spaces and complex numbers while elements of linear spaces over a finite ring or field and 
elements of the corresponding ring or field will be denoted without a "tilde".

Let $(\tilde e_1,\tilde e_2,...)$ be a basis in $H$. This
means that $\tilde x$ can be represented as
\begin{equation}
\tilde x =\tilde c_1 \tilde e_1+\tilde c_2 \tilde e_2+...
\label{G2}
\end{equation}
where $(\tilde c_1,\tilde c_2,...)$ are complex numbers.
It is assumed that there exists a complete set of commuting selfadjoint
operators $(\tilde A_1,\tilde A_2,...)$ in $H$ such that
each $\tilde e_i$ is the eigenvector of all these operators:
$\tilde A_j\tilde e_i ={\tilde \lambda}_{ji}\tilde e_i$. Then the
elements $(\tilde e_1,\tilde e_2,...)$ are mutually orthogonal:
$(\tilde e_i,\tilde e_j)=0$ if $i\neq j$ where (...,...) is
the scalar product in $H$. In that case the coefficients can
be calculated as
\begin{equation}
\tilde c_i = \frac{(\tilde e_i,\tilde x)}{(\tilde e_i,\tilde e_i)}
\label{G3}
\end{equation}
Their meaning is that
$|\tilde c_i|^2(\tilde e_i,\tilde e_i)/(\tilde x,\tilde x)$
represents the probability to find $\tilde x$ in the state
$\tilde e_i$. In particular, when $\tilde x$ and the basis
elements are normalized to one, the probability equals $|\tilde c_i|^2$.

In finite mathematics we can consider complex analogs of finite rings or fields. For example, one can consider the rings 
$R_{p^2}=R_p+iR_p$ or fields $F_{p^2}=F_p+iF_p$. The latter definition is valid if $p$ is prime and $p=3\,\,(mod\,\, 4)$
but quadratic extensions of $F_p$ can be also used if $p=1\,\,(mod\,\, 4)$ (see e.g. standard textbooks in Ref. \cite{textbooks}). We can now extend the function $f$ in Subsec. \ref{proof} such that it is now a function from
$R_{p^2}$ to $Z+iZ$ such that $f(a+ib)=f(a)+if(b)$. 

Therefore the transition from standard quantum theory to FQT can be performed as follows. We now describe quantum states not by
elements of complex Hilbert spaces but by elements of linear spaces $V$ over $R_{p^2}$ or $F_{p^2}$. This
means that elements $x\in V$ can be represented as
\begin{equation}
x =c_1 e_1+c_2e_2+...c_ne_n
\label{G2B}
\end{equation}
where the $c_j$ ($j=1,2,...n$) are elements of $R_{p^2}$ or $F_{p^2}$ and the $e_j$ ($j=1,2,...n$) are elements
of a basis in the given linear space over $R_{p^2}$ or $F_{p^2}$.

Since complex conjugation is the automorphism of $R_{p^2}$ and $F_{p^2}$ then, by analogy with conventional quantum theory, 
in FQT it is possible to formally consider situations when
linear spaces over $R_{p^2}$ or $F_{p^2}$ are supplied by a scalar product 
and it is also possible to consider analogs of  Hermitian operators.

Let us note that, in contrast to the situation with standard quantum theory, FQT
involves only finite dimensional spaces. In the case of one elementary particle this follows from the Zassenhaus theorem
\cite{Zassenhaus} that any IR over a field with finite characteristics can be only finite dimensional. Therefore any
elementary particles can have only a finite number of states. In the case of many particles the representation
space is the tensor product of single-particle spaces and therefore, for systems with a finite number
of elementary particles the representation space is finite dimensional too. 

One of the fundamental principle of physics is the correspondence principle which implies that for some situations
any new theory should reproduce results of the old well tested theory with a high accuracy. In the given case the
correspondence between the elements in Eqs. (\ref{G2}) and (\ref{G2B}) will take place 
if the absolute values for all the $c_j$ are much less than $p$, $f(c_j)={\tilde c_n}$ and similar is true for  
$c_{jk}=(e_j,e_k)$. Analogously one can introduce a correspondence between the operators $A$ in $V$ and
${\tilde A}$ in $H$.

Summarizing this discussion, we conclude that if $p$
is large then there exists a correspondence between the
description of physical states on the language of Hilbert spaces and self-adjoint operators
in them on one hand, and on the language of linear spaces over
$R_{p^2}$ and Hermitian operators in them on the other. However, in FQT probabilistic interpretation can be only 
approximate: it is valid only for
states the norm of which is much less than $p$.

From mathematical point of view, standard quantum theory can be treated as a theory of
representations of special real Lie algebras in complex Hilbert spaces while in FQT representation spaces 
are over a finite field or ring with characteristic $p$. As shown in Refs. \cite{lev4,monograph} and Sec. \ref{proof},  in the formal limit
$p\to\infty$ FQT recovers predictions of standard continuous theory.
Therefore classical mathematics describes many experiments with a high accuracy as a
consequence of the fact that the number $p$ is very large. However, since classical mathematics has
foundational problems by its own nature (as follows, for example, from G\"{o}del's incompleteness theorems),
ultimate quantum theory cannot be based on classical mathematics.

The above discussion indicates that, from the point of view of describing
quantum states, the Hilbert spaces in standard quantum theory contain a big redundancy of elements, 
and rational and real numbers play only the auxiliary role. On the contrary, in FQT  the description of 
states does not contain redundant elements, i.e. this description is much more thrifty than in standard quantum theory.

In view of the above discussion it seems obvious that fundamental quantum theory should be based on finite 
mathematics. However, in view of the circumstances described above, only a very few scientists work in this direction.
In particular, important results have been obtained by Planat, Saniga, Vourdas and others \cite{Saniga},
and a more detailed list of references can be found in Ref. \cite{monograph}.

In physics $p$ is a standard notation for the momentum but in number theory it is a
standard notation for the characteristic of a ring or field. In what follows it will be obvious in
what context the notation $p$ is used.

\section{Problem of space-time in quantum theory}
\label{intro}

Although quantum theory exists for more than 90 years, the problem of its
foundation is still widely debated. Although it is now obvious that physical intuition based on classical physics usually does not work for explaining quantum phenomena,
quantum theory inherited several important notions from classical one. 

For example, a rather strange feature of fundamental quantum theories (QED, QCD and electroweak theory) is that their derivation is based on local space-time Lagrangians
but the final formulation involves only the S-matrix in momentum representation and 
space-time is not present in this formulation at all. This is in the spirit of the Heisenberg
S-matrix program where description of quantum states at each moment of time $t$ is treated as unphysical and only the description of evolution  from the infinite past
when $t\to -\infty$ to the distant future when $t\to +\infty$ has a physical meaning.

In many cases fundamental quantum theories give impressive agreements with experiment but nevertheless the problem of substantiation of those theories remains open. The main inconsistency of the theories is that they contain divergent expressions for the S-matrix elements. The main reason is that the  Lagrangian densities contain products of local operator
fields at the same space-time points. As explained even in textbooks on local quantum field theories 
(see e.g. Ref. \cite{Bogolubov}), interacting local quantum fields
can be treated only as operator distributions, and a known fact from the theory of distributions is that their product at the same points is not a correct mathematical operation.  One of ideas of the string theory is that if products of fields at the same points (zero-dimensional objects) are replaced by products where the arguments of the 
fields belong to strings (one-dimensional objects) then there is hope that infinities will be less singular. However, a similar mathematical inconsistency exists in string theory as well and here the problem of infinities has not been solved yet.

A usual justification of the presence of such products is that they are needed to preserve locality. However, this argument is not consistent for the following reason.
Although the construction of the local quantized field $\psi(x)$ (where $x$ is a point in Minkowski space) is based on a single-particle field, the quantized field
is an operator in the Fock space for a system with an infinite number of particles, and the argument $x$ does not refer 
to any particle. It is not an eigenvalue of the position operator and it is only
an integration parameter for the full Lagrangian. 

Probably the "strongest" justification is that the mentality of the majority of physicists is that agreement with experiment is much more important than mathematical rigor. For them it does not ring a bell that the existence of mathematical inconsistencies
in standard quantum theory might be an indication that the theory is not universal. A historical analogy is that
classical mechanics describes many data with high accuracy but fails when $v/c$ is not small.

Let us note that even in classical mechanics particle coordinates and time can be treated in different ways.
A standard treatment of this theory is that its goal is to solve equations of motion and get
classical trajectories where coordinates and momenta are functions of time $t$. In 
Hamiltonian mechanics the action can be written as $S = S_0-\int Hdt$ where $H$ is the Hamiltonian, $S_0$ does not 
depend on $t$ and is called the abbreviated action. Suppose now that
one wishes to consider a problem which is usually treated as less general: to find
not the dependence of the coordinates and momenta on $t$ but only possible forms of
trajectories in the phase space without mentioning time at all. If the energy is
a conserved physical quantity then, as described in textbooks, this problem
can be solved by using the Maupertuis principle involving only $S_0$.

However, the latter problem {\it is not} less general than the former one. For
illustration we consider the one-body case. Suppose that by using the Maupertuis
principle one has solved the problem with some initial values of coordinates
and momenta. Let $s$ be a parameter characterizing the particle trajectory, i.e. the particle radius-vector ${\bf r}$,
the momentum ${\bf p}$ and the energy $E$ are functions of $s$.  The particle velocity ${\bf v}$ in units $c=1$ 
{\it is defined} as ${\bf v}(s)={\bf p}(s)/E(s)$. At this stage the problem does not contain $t$ yet. One can {\it define}
$t$ by the condition that $dt=|d{\bf r}|/|{\bf v}|$ and
hence the value of $t$ at any point of the trajectory can be obtained by integration. Hence 
the general problem of classical mechanics can be initially formulated without mentioning $t$ while if
for some reasons one prefers to work with $t$ then its value can flow only in the positive direction since $dt>0$.

Another point of view is that, at least on classical level, time is a primary quantity while the coordinates ${\bf r}$ 
of each free particle should be {\it defined} in terms of momenta and time as 
\begin{equation}
d{\bf r}={\bf v}dt=\frac{{\bf p}}{E}dt
\label{coordmom}
\end{equation}
where $E=(m^2+{\bf p}^2)^{1/2}$ and $m$ is the particle mass.
Such a definition of coordinates
is similar to that in General Relativity (GR) where distances are defined in terms of time needed for light to travel from one point to
another.

On quantum level the treatment of particle coordinates and time becomes much more complicated. The postulate of quantum theory
is that for any physical quantity there should exist a corresponding selfadjoined operator. As noted by Pauli (see p. 63 of Ref. \cite{Pauli}), 
at early stages of quantum theory some
authors treated time $t$ as an operator commuting with $H$ as $[H,t] = i\hbar$, i.e. $H$ and $t$ are canonically conjugated. 
However, there are several reasons why such a treatment is not correct. For example (see e.g. Ref. \cite{Leon}), the conjugated operators
should necessarily have the same spectrum, time has the continuous spectrum in the range $(-\infty,+\infty)$ while
the Hamiltonian is usually bounded below and a part of its spectrum may be discrete. 

It is usually assumed that in quantum theory the quantity
$t$ can be only a classical parameter describing evolution of a quantum system by the time dependent 
Schr\"{o}dinger equation. The usual justification is that in the formal limit $\hbar\to 0$ it
becomes the Hamilton-Jacobi equation. Moreover, the justification of standard choice for
different operators (e.g. coordinate, momentum, angular momentum operators and others) is that such a choice has a correct classical limit. However,  the correct classical limit does not guarantee the correct behavior on quantum level.
For example, if $A$ and $B$ are two operators such that $B$ becomes zero in classical limit then the operators
$A$ and $A+B$ have the same classical limit but on quantum level they may have considerably different 
properties.

A problem arises why the principle of
quantum theory that every physical quantity is defined by an operator does not apply
to time. In the literature the problem of time is also often formulated such that "the time of GR and of ordinary Quantum Theory are mutually incompatible notions" (see e.g. Ref. \cite{Anderson}). 
As noted by several authors, (see e.g. Refs. \cite{RovelliBook,RovelliTime,Keaton}), $t$ cannot be treated as a fundamental
physical quantity. The reason is that all fundamental physical
laws do not require time and the quantity $t$ is obsolete on fundamental level. 

In quantum theory a problem arises "how  to forget time" (by analogy with the Maupertuis principle), construct a
theory (in particular quantum gravity) which does not involve time at all and in what approximations classical time can be reconstructed. This is a very complicated problem which has been discussed in detail in Refs. \cite{RovelliBook,RovelliTime}.

One can also consider a situation when a quantum system under
consideration is a small subsystem of a big system where the big subsystem - the environment, is
strongly classical. Then one can define $t$ for the environment as described
above. The author of Ref. \cite{Keaton} considers a scenario when the system as a whole is
described by the stationary Schr\"{o}dinger equation $H\Psi	 = E\Psi$	 but the small quantum
subsystem is described by the time dependent Schr\"{o}dinger equation where $t$ is
defined for the environment as $t=\partial S_0/\partial E$. In this scenario it is clear why a quantum system is described by the  
Schr\"{o}dinger equation depending on the classical parameter $t$ which is not an operator: because 
$t$ is the physical quantity
characterizing not the quantum system but the environment. This scenario seems also natural because
it is in the spirit of the Copenhagen interpretation of quantum theory: the evolution of a quantum system
can be characterized only in terms of measurements which in the Copenhagen interpretation are treated as
interactions with classical objects. However, this scenario encounters several problems. 
For example, the environment can be a classical
object only in some approximation and, as noted in Ref. \cite{Keaton}, the above scenario does not solve the problem of quantum jumps. 

The authors of Ref. \cite{Leon} state that the Pauli objection can be circumvented if 
 one uses an external system  to track time, so that "time arises as correlations between the system
and the clock". In this case, the time operator can be defined. It is not
conjugate to the system Hamiltonian, but its eigenvalues still satisfy the Schr\"{o}dinger equation for
arbitrary Hamiltonians. Such an approach is to some extent in the spirit of Ref. \cite{Keaton}. The authors of Ref. \cite{Leon} refer to the extensive literature where the time operator has been discussed. In any case,
the problem to deal or not with the time operator depends on the 
physical situation and there is no universal choice of the time operator which follows from first principles of quantum theory. 

In contrast to time, it is usually believed that in quantum theory the position operator has a clear physical meaning. For example, in nonrelativistic quantum mechanics
the position and momentum operators are related to each other by the Fourier transform. As a consequence, we have the famous Heisenberg uncertainty relations or {\it vice versa},
from these relations it follows that the operators are related to each other by the Fourier transform. Many authors (including Heisenberg, Dirac and others) gave different
arguments in favor of such relations. The great success of the early quantum theory was that the nonrelativistic 
Schr\"{o}dinger equation gives a good description of the hydrogen energy levels and the Dirac equation gives a good description of the fine structure of those levels in the approximation $(v/c)^2$.

However, from the point of view of the present knowledge, 
 the Schr\"{o}dinger and Dirac equations should be treated as follows.
 As follows from Feynman diagrams for the one-photon exchange, in the approximation 
up to  $(v/c)^2$  the electron in the hydrogen atom can be described in the potential formalism where the potential acts on the wave function (WF) in momentum space.  So for calculating energy levels one should solve the eigenvalue problem for the Hamiltonian with this potential. This is an integral equation which can be solved by different methods. One of the convenient methods is to apply the Fourier transform and get standard Schr\"{o}dinger or Dirac equation in coordinate representation with the Coulomb potential. Hence the fact that the results for energy levels are in good agreement with experiment shows that QED defines the potential correctly and {\it standard coordinate 
Schr\"{o}dinger and Dirac equations are only convenient mathematical ways of solving the eigenvalue problem in the approximation up to $(v/c)^2$}. For this problem 
the physical meaning of the position operator is not important at all. One can consider other transformations of the original integral equation and define other position operators. The fact that for non-standard choices one might obtain something different from the Coulomb potential is not important on quantum level. 

The Schr\"{o}dinger and Dirac equations work with a high accuracy because the fine structure constant
$\alpha$ is small and, as a consequence, the effects beyond the single-particle approximation (e.g. the Lamb
shift) are small. However, consider a hypothetical situation where a Universe is such that the value of $\alpha$ is of the
order of unity or greater. Although it is not known (even if $\alpha$ is small) whether the perturbation series of QED 
converges or not, the logical structure of QED remains the same. At the same time, the single-particle approximation
is not valid anymore and the Schr\"{o}dinger and 
Dirac equations do not define the hydrogen energy levels even approximately. In other words, in this
situation the application of those equations for calculating the hydrogen energy level does not have a
physical meaning.  

The fact that in our world  the Schr\"{o}dinger and Dirac equations describe the hydrogen energy
level with a high accuracy, is usually treated as a strong argument that the coordinate and momentum representations
should be related to each other by the Fourier transform.  However, as follows from the above considerations,
this fact takes place only because we are lucky that the value of $\alpha$ in our
Universe is small. 

Historically the great success of the Dirac equation was that the existence of negative energy levels was interpreted
as existence of antiparticles. However, as shown by Pauli \cite{Pauli2}, local quantum fields do not have probabilistic interpretation
because in the case of fields with an integer spin there is no invariant subspace where the spectrum of the charge operator has
a definite sign while in the case of fields with a half-integer spin there is no invariant subspace
where the spectrum of the energy operator has a definite sign. The absence of probabilistic interpretation 
follows also from the fact that representations
of the Poincare group describing local fields are nonunitary because they are induced from nonunitary
representations of the Lorentz group. 
It is now clear that antiparticles exist simply because for
any IR of the Poincare algebra with positive energies there exists a corresponding IR with
negative energies. In QFT the fact that the masses of a particle and its antiparticle are the same follows
from the CPT theorem for free local theories. However, in FQT this fact can be proved without any 
assumption about locality \cite{monograph}.

As shown by Newton and Wigner \cite{NW}, relativistic position operator differs from the nonrelativistic one
but the basic feature that the momentum and position operators are related to each other by the Fourier
transform remains in the Newton-Wigner construction as well. This postulate is a good illustration of the fact mentioned at the beginning of this section that quantum theory inherited many its features from the classical one. 
The relation between the coordinates and momenta is analogous to one between the coordinates and wave vectors in classical electrodynamics. A known effect here is the wave packet spreading (WPS). In classical electrodynamics the wave packet consists of many particles but in quantum theory the effect takes place even for a single-particle WF. 

At the very beginning of quantum theory several physicists (e.g. de Broglie) argued that the WPS effect should not take place in quantum theory and the single particle should not be described by the time dependent Schr\"{o}dinger equation. On the other hand, as shown by Darwin \cite{Darwin}, for macroscopic particles the WPS effect is negligible. 
It is also believed that in experiments with atoms and elementary particles the time is so small that the WPS effect does not manifest itself. Probably for those reasons the majority of physicists do not treat the WPS effect as a drawback of the theory.

However, photons from distant stars can travel to Earth even for billions of years and for them the WPS effect cannot be neglected. As shown in Ref. \cite{lev1}, the WPS effect for such photons results in a fundamental quantal paradox that predictions of the theory contradict our experience on how we observe stars. The paradox can be resolved if the position operator is essentially different from standard one and the coordinate and momentum representations are not related by the Fourier transform. 

One can discuss different choices of the position operator but in any case the choice is not dictated by first principles of quantum theory. History of physics tells us that in any theory it is desirable to have the least possible amount of notions. Quantum theory is more general than classical one and so at some conditions it should reproduce all the results of classical theory including classical equations of motion. However, it does not mean that quantum theory should explicitly involve particle coordinates and time.

The results of the paper related to the nature of time are described in Sec. \ref{classicalFQT}. Here it is shown that there exist scenarios when
classical equations of motion can be obtained from quantum theory without using any classical notions such as 
coordinates, time, position operator, standard semiclassical approximation etc. The goal of Secs. \ref{2particles} - \ref{S9}
 is to prepare the reader for 
understanding these results. Here the consideration is based on the results obtained in our previous publications,
mainly in Refs. \cite{lev4,JPA,monograph}. Those results have been obtained with extensive calculations but in this
paper we explain the meaning of the results and argue that they are very natural.

\section{Symmetry on quantum level}
\label{symmetry}

In relativistic quantum theory the usual approach to symmetry on quantum level follows. 
Since the Poincare group is the group of motions of Minkowski space, quantum states should be described by representations of this group. 
This implies that the representation generators commute according to the commutation relations of the Poincare group Lie algebra:
\begin{eqnarray}
&&[P^{\mu},P^{\nu}]=0,\quad [P^{\mu},M^{\nu\rho}]=-i(\eta^{\mu\rho}P^{\nu}-\eta^{\mu\nu}P^{\rho})\nonumber\\
&&[M^{\mu\nu},M^{\rho\sigma}]=-2i (\eta^{\mu\rho}M^{\nu\sigma}+\eta^{\nu\sigma}M^{\mu\rho}-
\eta^{\mu\sigma}M^{\nu\rho}-\eta^{\nu\rho}M^{\mu\sigma})
\label{PCR}
\end{eqnarray}
where $\mu,\nu=0,1,2,3$, $P^{\mu}$ are the operators of the four-momentum and  $M^{\mu\nu}$ are the operators of
Lorentz angular momenta. This approach is in the spirit of Klein's Erlangen program in mathematics.
However, as we argue in Refs. \cite{monograph,PRD} and Sec. \ref{intro}, quantum theory should not be based on 
classical space-time background
and the approach should be the opposite. Each system is described by a set of independent operators.
By definition, the rules how they commute with each other define the symmetry algebra. 
In particular, {\it by definition}, Poincare symmetry on quantum level means that the operators commute
according to Eq. (\ref{PCR}). This definition does not involve Minkowski space at all.

Such a definition of symmetry on quantum level has been explained to me by Leonid Avksent'evich Kondratyuk during our
collaboration. I believe that this replacement of the standard paradigm is fundamental for understanding 
quantum theory, and I did not succeed in finding a similar idea in the literature. This idea is to some extent
in the spirit of Ref. \cite{Dir}. Here Dirac proposed different forms of relativistic dynamics  which are defined by choosing which operators in Eq. (\ref{PCR}) are free and which of them are interaction dependent. 

Analogously, the definition of dS and AdS symmetries on quantum level should not
involve the fact that the dS and AdS groups are the groups of motions of dS and AdS spaces, respectively.
Instead, {\it the definition} is that the operators $M^{ab}$ ($a,b=0,1,2,3,4$, $M^{ab}=-M^{ba}$)
describing the system under consideration satisfy the
commutation relations (\ref{CR}).

With such a definition of symmetry on quantum level, dS and AdS
symmetries look more natural than Poincare symmetry. In the
dS and AdS cases all the ten representation operators of the symmetry
algebra are angular momenta while in the Poincare case only six
of them are angular momenta and the remaining four operators
represent standard energy and momentum. If we {\it define} the
operators $P^{\mu}$ as $P^{\mu}=M^{4\mu}/(2R)$ where $R$ is a parameter with the dimension
$length$ then in the formal
limit when $R\to\infty$, $M^{4\mu}\to\infty$ but the quantities
$P^{\mu}$ are finite, Eqs. (\ref{CR}) become Eqs. (\ref{PCR}). This procedure is called contraction and 
in the given case it is the same for the dS or AdS symmetries.

In the literature, Poincare, dS and AdS symmetries are usually associated not with the
corresponding algebras but (in the spirit of the Erlangen program) with the background space
invariant under the action of the corresponding group.
Those spaces are characterized by the curvature called the cosmological constant $\Lambda$ (CC)
such that $\Lambda=0$, $\Lambda>0$ and $\Lambda<0$ respectively.
The expressions for $\Lambda$ in terms of $R$ are
$\Lambda=0$, $\Lambda=3/R^2$ and $\Lambda=-3/R^2$, respectively.

It is obvious that FQT can involve only numbers and cannot contain any dimensionful quantities. Equations
(\ref{CR}) contain no parameters and it is often said that those expressions are written in units $\hbar/2=c=1$.
This phrase might create a wrong impression that expressions with $\hbar$ and $c$ are primary while Eqs.
(\ref{CR}) are secondary, but, as noted in Sec. \ref{fundamentaltheories} the situation is the opposite. 

Let us now define the notion of elementary particle. 
Although theory of elementary particles exists for a rather long period of time, there is no commonly accepted definition of elementary particle in this theory. In the spirit of the above definition of symmetry on quantum level and 
Wigner's approach to Poincare symmetry
\cite{Wigner}, a general definition, not depending on the choice of the classical background 
and on whether we consider a local or nonlocal theory, is  that a particle is elementary if 
the set of its WFs is the space of an IR of the symmetry algebra in the given theory. 

The explicit construction of IRs of the dS and AdS algebras describing elementary particles (see e.g. Refs.
\cite{lev4,monograph}) shows that it is possible to find a basis  where the spectrum of all the representation
operators is discrete. Therefore such IRs can be used in both, standard theory and FQT. At the same time,
for IRs  describing elementary particles in Poincare invariant theory the spectrum of some operators is
necessarily continuous. Therefore such IRs cannot be used in FQT.

By definition, the tensor product of IRs
corresponding to $N$ particles describes a system where those particles are free. The representation
operators for the free $N$-particle systems are sums of the corresponding single-particle operators.
In the present paper we consider only systems of free particles, i.e. there is no interaction between the
particles. A problem arises whether the cosmological repulsion and gravity are not interactions but simply 
(kinematic) consequences of dS/AdS symmetry and/or FQT (i.e. the fact that $p$ is finite), i.e. from the 
formal point of they can take place even in systems of free particles.

In standard nonrelativistic approximation, 
gravity is characterized by the term $-Gm_1m_2/r$ in the mean value of the 
two-particle mass operator.
Here $m_1$ and $m_2$
are the particle masses and $r$ is the distance between
the particles. Since the kinetic energy is always positive,
the free nonrelativistic mass operator is positive definite
and therefore there is no way to obtain gravity in the
framework of the free theory. Analogously, in Poincare
invariant theory the spectrum of the free two-particle mass
operator belongs to the interval $[m_1+m_2,\infty )$ while the existence of gravity necessarily requires
that the spectrum should contain values less than $m_1+m_2$.

In theories where the symmetry algebra is the AdS algebra so(2,3), the structure of IRs
is well-known (see e.g. Refs. \cite{lev4,monograph}). In particular, for positive
energy IRs the AdS Hamiltonian has the spectrum in
the interval $[m,\infty )$ and $m$ has the meaning 
of the AdS mass. Therefore the situation is pretty much
analogous to that in Poincare invariant theories.
In particular, the free two-particle mass operator again
has the spectrum in the interval $[m_1+m_2,\infty )$
and therefore there is no way to reproduce gravitational
effects in the free AdS theory.

In contrast to the situation in Poincare and AdS theories, the free two-particle mass operator in dS theory
is not bounded below by the value of $m_1+m_2$. The results of Ref. \cite{dS,JPA,monograph} show that
this property by no means implies that the theory is unphysical.  In addition, the existing experimental data (see e.g. Ref. \cite{Perlmutter}) practically exclude the possibility that $\Lambda \leq 0$. As shown in Refs. \cite{dS,JPA}
(see also the next section) the cosmological repulsion naturally arises in free systems described in the framework
of the dS theory. Therefore if one has a choice between Poincare, AdS
and dS symmetries then the only chance to describe the cosmological repulsion and gravity in a free theory is to choose dS symmetry.

\section{A system of two particles in standard quantum dS theory}
\label{2particles}

As shown in Ref. \cite{JPA}, by using the results of the book \cite{Mensky} on IRs of the dS group one can explicitly 
construct IRs of the dS algebra describing elementary particles. In this paper we are interested not in 
elementary particles but in macroscopic bodies. If we consider systems of particles such that the distances
between them are much greater than their sizes then the internal structure of the particles is not important and
it suffices to describe each particle only by the variables characterizing the motion of each particle as a
whole. The WFs describing such a motion are the same as for elementary particles i.e. we can use IRs of the
dS algebra. Since spin is a pure quantum notion which disappears in classical limit, we will consider only 
spinless IRs and will not consider massless and tachyon representations.

In Poincare theory any massive IR can be implemented in the Hilbert space of functions $\chi({\bf v})$ on
the Lorenz 4-velocity hyperboloid with the points $v=(v_0,{\bf v}),\,\, v_0=(1+{\bf v}^2)^{1/2}$ such that 
$\int\nolimits |\chi({\bf v})|^2d\rho({\bf v}) <\infty$ and $d\rho({\bf v})=d^3{\bf v}/v_0$ is the Lorenz
invariant volume element. For positive and negative energy IRs the value of energy is $E=\pm mv_0$
respectively where $m$ is the particle mass {\it defined as the positive square root} $(E^2-{\bf P}^2)^{1/2}$.
Therefore for massive IRs, $m>0$ by definition.

It is usually assumed that the energy for real particles should be positive. However, the choice of the energy sign is only
the matter of convention but not the matter of principle. It is only important that the energy sign for all the
particles in question is the same because otherwise the conservation of energy will not take place.
In the literature the positive energy IRs are usually associated with particles and  the negative energy IRs --- with
the corresponding antiparticles. Then after the second quantization the energies of both, particles and
antiparticle become positive.

In contrast to Poincare theory, IRs in dS theory can be implemented only on two Lorenz hyperboloids,
i.e. the Hilbert space for such IRs consist of sets of two functions $(\chi_1({\bf v}),\chi_2({\bf v}))$ such that 
$$\int\nolimits (|\chi_1({\bf v})|^2+|\chi_2({\bf v})|^2)d\rho({\bf v}) <\infty$$
In Poincare limit one dS IR splits into two IRs of the Poincare algebra with positive and negative energies.
In Ref. \cite{JPA} we argue that this implies that one IR of the dS algebra describes a particle and its
antiparticle simultaneously. Since in the present paper we do not deal with antiparticles, we give
only expressions for the action of the operators on the upper hyperboloid \cite{JPA}:
\begin{eqnarray}
&&{\bf M}=2l({\bf v}),\quad {\bf N}=-2i v_0 \frac{\partial}{\partial {\bf v}},\quad {\bf B}=m_{dS} {\bf v}+2i [\frac{\partial}{\partial {\bf v}}+
2{\bf v}({\bf v}\frac{\partial}{\partial {\bf v}})+\frac{3}{2}{\bf v}]\nonumber\\
&& {\cal E}=m_{dS} v_0+2i v_0({\bf v}
\frac{\partial}{\partial {\bf v}}+\frac{3}{2})
\label{IR1}
\end{eqnarray}
where ${\bf M}=\{M^{23},M^{31},M^{12}\}$, ${\bf
N}=\{M^{01},M^{02},M^{03}\}$, ${\bf
B}=\{M^{41},M^{42},M^{43}\}$, ${\bf l}({\bf v})=-i{\bf v}\times \partial/\partial {\bf
v}$, ${\cal E}=M^{40}$ and $m_{dS}$ is a positive quantity. 

This implementation of the IR is convenient for the transition to Poincare limit. Indeed, the operators
of the Lorenz algebra in Eq. (\ref{IR1}) are the same as in the IR of the Poincare algebra. Suppose
that the limit of $m_{dS}/(2R)$ when $R\to\infty$ is finite and denote this limit as $m$. Then in the
limit $R\to\infty$ we get standard expressions for the operators of the IR of the Poincare algebra
where $m$ is standard mass, $E={\cal E}/(2R)=mv_0$ and ${\bf P}={\bf B}/(2R)=m{\bf v}$. For this
reason $m_{dS}$ has the meaning of the dS mass. In contrast to $m$, $m_{dS}$ is dimensionless.
Since Poincare symmetry is a special case of dS one, $m_{dS}$ is more fundamental than $m$.
Since Poincare symmetry works with a high accuracy, the value of $R$ is supposed to be very large.
Then even dS masses of elementary particles are very large. 

For example, according to Ref. \cite{Perlmutter},
$R\approx 10^{26}meters$. The conclusion of this work on $R$ is based not on the consideration of the dS
algebra but from the fit to the Friedman-Robertson-Walker model. This value of $R$ is in the spirit
of modern cosmology that the Universe has approximately the same size. However, the model depends 
on parameters
and therefore the validity of the conclusion cannot be accepted for granted. In particular, the value of $R$
may be much greater than $10^{26}meters$. However, even for this value of $R$
the dS masses of the electron,
the Earth and the Sun are of the order of $10^{39}$, $10^{93}$ and $10^{99}$, respectively. 
The fact that even the dS mass of the electron is very large poses a question whether the electron
is a true elementary particle.

Consider the non-relativistic approximation when $|{\bf v}|\ll
1$. If we wish to work with units where the dimension of
velocity is $meter/sec$, we should replace ${\bf v}$ by ${\bf
v}/c$. If ${\bf p}=m{\bf v}$ then it is clear from the
expression for ${\bf B}$ in Eq. (\ref{IR1}) that ${\bf p}$ becomes the real momentum ${\bf P}$
only in the limit $R\to\infty$. At this
stage we do not have any coordinate space yet. However, if we
assume that semiclassical approximation is valid, then, by analogy with
standard quantum mechanics, we can {\it define} the position
operator ${\bf r}$ as $i\partial/\partial {\bf p}$. 

In classical approximation we can treat
${\bf p}$ and ${\bf r}$ as usual vectors. Then as follows from Eq. (\ref{IR1})
\begin{equation}
{\bf P}= {\bf p}+mc{\bf r}/R, \quad H = {\bf p}^2/2m +c{\bf p}{\bf r}/R,\quad {\bf N}=-m{\bf r}
\label{PH}
\end{equation}
where $H=E-mc^2$ is the classical nonrelativistic Hamiltonian. As follows from these expressions, 
\begin{equation}
H({\bf P},{\bf r})=\frac{{\bf P}^2}{2m}-\frac{mc^2{\bf r}^2}{2R^2}
\label{HP}
\end{equation}

The last term in Eq. (\ref{HP}) is the dS correction to
the non-relativistic Hamiltonian. It is interesting to note
that the non-relativistic Hamiltonian depends on $c$ although
it is usually believed that $c$ can be present only in
relativistic theory. This illustrates the fact mentioned in
Sec. \ref{symmetry} that the transition to nonrelativistic theory
understood as $|{\bf v}|\ll 1$ is more physical than that
understood as $c\to\infty$. The presence of $c$ in Eq.
(\ref{HP}) is a consequence of the fact that this expression is
written in standard units. In nonrelativistic theory $c$ is
usually treated as a very large quantity. Nevertheless, the
last term in Eq. (\ref{HP}) is not large since we assume
that $R$ is very large.

As follows from Eq. (\ref{HP}) and the Hamilton equations, in dS theory a free particle
moves with the acceleration given by
\begin{equation}
{\bf a}={\bf r}c^2/R^2
\label{accel}
\end{equation}
where ${\bf a}$ and ${\bf r}$ are the acceleration and the radius vector of the particle, respectively.
Since $R$ is very large, the acceleration is not negligible only at cosmological distances
when $|{\bf r}|$ is of the order of $R$. The result (\ref{accel}) can be obtained not only from
Hamilton equations but by different ways. For example, assuming that the Hamiltonian is a
conserved physical quantity, this result can be obtained from the Maupertuis principle or from
Eq. (\ref{coordmom}) as noted in Sec. \ref{intro}.

Let us now consider whether the result (\ref{accel}) is compatible with GR. 
The dS space is a four-dimensional manifold in the five-dimensional space defined by
\begin{equation}
x_1^2+x_2^2+x_3^2+x_4^2-x_0^2=R^2
\label{dSspace}
\end{equation}
In the formal limit $R\to\infty$ the action of the dS group in
a vicinity of the point $(0,0,0,0,x_4=R)$ becomes the action
of the Poincare group on Minkowski space. The dS space 
can be parameterized without using
the quantity $R$ at all if instead of $x_a$ ($a=0,1,2,3,4$) we
define dimensionless variables $\xi_a=x_a/R$. It is also clear
that  the elements of the SO(1,4) group do not depend on $R$
since they are products of conventional and hyperbolic
rotations. So the dimensionful value of $R$ appears only if one
wishes to measure coordinates on the dS space in terms of
coordinates of the flat five-dimensional space where the dS
space is embedded in. This requirement does not have a
fundamental physical meaning. Therefore the value of $R$
defines only a scale factor for measuring coordinates in the dS
space. 

With the parameterization of dS space as in Eq. (\ref{dSspace}) 
the metric tensor on this space is
\begin{equation}
g_{\mu\nu}=\eta_{\mu\nu}-x_{\mu}x_{\nu}/(R^2+x_{\rho}x^{\rho})
\label{metric}
\end{equation}
where $\mu,\nu,\rho = 0,1,2,3$, $\eta_{\mu\nu}$ is the Minkowski metric tensor,  
and a summation
over repeated indices is assumed. It is easy to calculate the
Christoffel symbols in the approximation where all the
components of the vector $x$ are much less than $R$:
$\Gamma_{\mu,\nu\rho}=-x_{\mu}\eta_{\nu\rho}/R^2$. Then a
direct calculation shows that in the nonrelativistic
approximation the equation of motion for a single particle is
 the same as in Eq. (\ref{accel}).

Another way to show that Eq. (\ref{accel}) is compatible with GR follows. The known result of GR is that if the metric
is stationary and differs slightly from the Minkowskian one
then in the nonrelativistic approximation the curved space-time
can be effectively described by a gravitational potential
$\varphi({\bf r})=(g_{00}({\bf r})-1)/2c^2$. We now express
$x_0$ in Eq. (\ref{dSspace}) in terms of a new variable $t$ as
$x_0=t+t^3/6R^2-t{\bf x}^2/2R^2$. Then the expression for the
interval becomes
\begin{equation}
ds^2=dt^2(1-{\bf r}^2/R^2)-d{\bf r}^2-
({\bf r}d{\bf r}/R)^2
\label{II67}
\end{equation}
Therefore, the metric becomes stationary and $\varphi({\bf r})=-{\bf r}^2/2R^2$ in agreement with Eq. (\ref{accel}).

Consider now a system of two free particles in dS space. Let $({\bf r}_i,{\bf a}_i)$
$(i=1,2)$ be their radius vectors and accelerations, respectively. Then Eq. (\ref{accel}) is
valid for each particle if $({\bf r},{\bf a})$ is replaced by $({\bf r}_i,{\bf a}_i)$, respectively.
Now if we define the relative radius vector ${\bf r}={\bf r}_1-{\bf r}_2$ and the 
relative acceleration ${\bf a}={\bf a}_1-{\bf a}_2$ then they will satisfy the same Eq. (\ref{accel})
which shows that the dS antigravity is repulsive. It terms of $\Lambda$ it reads
${\bf a}=\Lambda{\bf r}c^2/3$ and therefore in the AdS case we have attraction rather than repulsion. 

Let us now consider a system of two free particles in the framework of the representation of the dS
algebra. The particles are described by the
variables ${\bf P}_j$ and ${\bf r}_j$ ($j=1,2$). Define standard nonrelativistic variables
\begin{eqnarray}
&&{\bf P}_{12}={\bf P}_1+{\bf P}_2,
\quad {\bf q}=(m_2{\bf P}_1-m_1{\bf P}_2)/(m_1+m_2)\nonumber\\
&&{\bf R}_{12}=(m_1{\bf r}_1+m_2{\bf r}_2)/(m_1+m_2),\quad
{\bf r}={\bf r}_1-{\bf r}_2
\label{2body}
\end{eqnarray}
Then, as follows from Eq. (\ref{PH}), in the
nonrelativistic approximation the two-particle quantities ${\bf P}$, ${\bf E}$ and ${\bf N}$ are given by
\begin{equation}
{\bf P}= {\bf P}_{12},\quad E = M+\frac{{\bf P}_{12}^2}{2M} -\frac{Mc^2{\bf R}_{12}^2}{2R^2},\quad {\bf N}=-M{\bf R}_{12}
\label{2PE}
\end{equation}
where
\begin{equation}
M = M({\bf q},{\bf r})=
m_1+m_2 +H_{nr}({\bf r},{\bf q}),\quad 
H_{nr}({\bf r},{\bf q})=\frac{{\bf q}^2}{2m_{12}}-\frac{m_{12}c^2{\bf r}^2}{2R^2}
\label{2M}
\end{equation}
and $m_{12}$ is the reduced two-particle mass. Here the operator $M$ acts in the space of functions
$\chi({\bf q})$ such that $\int |\chi({\bf q})|^2d^3{\bf q}<\infty$ and ${\bf r}$ acts in this space as
${\bf r}=i\partial/\partial {\bf q}$. 

It now follows from Eq. (\ref{IR1}) that $M$ has the meaning of the two-body mass. This can also
be shown \cite{JPA,monograph} from the fact that $M$ is the Casimir operator, i.e. it commutes with
all representation operators. As follows from Eq. (\ref{CR}), in the dS case the
Casimir operator of the second order is
\begin{eqnarray}
&&I_2 =-\frac{1}{2}\sum_{ab} M_{ab}M^{ab}={\cal E}^2+{\bf N}^2-{\bf B}^2-{\bf J}^2
\label{casimir}
\end{eqnarray}
According to
the known Schur lemma in representation theory, all elements in the space of IR are eigenvectors 
of the Casimir
operators with the same eigenvalue. A direct calculation shows that for the operators (\ref{IR1})
the numerical value of $I_2$ is $m_{dS}^2+9$. One can also show \cite{JPA} that for IRs
with spin $I_2=m_{dS}^2-{\bf s}^2+9$ where ${\bf s}$ is the spin operator. Then the explicit calculation
\cite{JPA} shows that for the two-body system $I_2=M^2-{\bf S}^2+9$ where ${\bf S}$ is the spin
operator for the two-body system, i.e. the angular momentum in the rest frame.
Therefore $M({\bf q},{\bf r})$ is the internal two-body Hamiltonian. Then, by analogy with the derivation of Eq. (\ref{accel}),
it can be shown in different ways that in semiclassical approximation the relative
acceleration is given by the same expression (\ref{accel}) but now
${\bf a}$ is the relative acceleration and ${\bf r}$ is the
relative radius vector.

The fact that two free particles have a relative acceleration
is known for cosmologists considering dS symmetry on
classical level. This effect is called the dS antigravity. The
term antigravity in this context means that the particles
repulse rather than attract each other. In the case of the dS
antigravity the relative acceleration of two free particles is
proportional (not inversely proportional!) to the distance
between them. As shown above, this classical result is a direct consequence of GR.

The experimental results obtained in 1998 (see e.g. Ref. \cite{Perlmutter}) is that $R$ is of the order of $10^{26}meters$, 
i.e. $\Lambda$ is very small but, as stated in Ref. \cite{Perlmutter}), the accuracy of the experiment is of the
order of 5\% and therefore the cases $\Lambda\leq 0$ are practically excluded. This created the following problem. 

In textbooks written before 1998 (when the cosmological
acceleration was discovered) it is often claimed that $\Lambda$ is not needed since its presence
contradicts the philosophy of GR: matter creates curvature of space-time, so  
in the absence of matter space-time should be flat (i.e. Minkowski) while empty dS space is not flat.
This philosophy has historical roots in view of the well-known fact that first Einstein introduced $\Lambda$
into his equations and then said that it was the greatest blunder of his life. The problem whether the empty
space-time may have a nonzero curvature was also discussed in the dispute between Einstein and de Sitter. 

However, such a philosophy has no physical meaning since the curvature is only a mathematical way to describe
the motion of real bodies and therefore the curvature does not have a physical
meaning for empty space-time. However, in view of the above statement, 
the discovery of the fact that $\Lambda \neq 0$ has ignited many discussions.
The most popular approach is as follows. One can move the term with $\Lambda$ in the Einstein equations from the
left-hand side to the right-hand one. 
Then the term with $\Lambda$ is treated as the stress-energy tensor of a hidden matter which is called
dark energy. With such an approach one implicitly
returns to Einstein's point of view that a curved space-time cannot
be empty. In other words, this is an
assumption that the Poincare symmetry is fundamental while the dS one
is emergent. With the observed value of $\Lambda$ this
dark energy contains approximately 75\% of the energy of the Universe. In this approach $G$ is treated as a
fundamental constant and one might try to express $\Lambda$ in terms of $G$. The existing quantum theory
of gravity cannot perform this calculation unambiguously since the theory contains strong divergences.
With a reasonable cutoff parameter, the result for $\Lambda$ is such that in units where $\hbar/2=c=1$,
$G\Lambda$ is of the order of unity. This result is expected from dimensionful considerations since in these units,
the dimension of $G$ is $length^2$ while the dimension of $\Lambda$ is $1/length^2$. However, this value
of $\Lambda$ is greater than the observed one by 122 orders of magnitude. 
In supergravity the disagreement can be reduced but even in best scenarios it exceeds 40 orders of 
magnitude. This problem is called the CC problem or dark energy problem.

Several authors criticized this approach from the following considerations. GR without the contribution of
$\Lambda$ has been confirmed with a good accuracy in experiments in Solar System. If $\Lambda$ is as
small as it has been observed then it can have a significant effect only at cosmological distances while
for experiments in Solar System the role of such a small value is negligible. The authors of Ref.
\cite{Bianchi} titled "Why All These Prejudices Against a Constant?", note that since the solution of the Einstein 
equations depends on two arbitrary constants $G$ and $\Lambda$
it is not clear why we should think that only a special case $\Lambda=0$ is allowed. 

In our approach the result for the cosmological acceleration has been obtained without using dS space, its
metric, connection etc. This result is simply a consequence of standard dS quantum mechanics of two free bodies
and the calculation does nor involve any geometry. The fact that $\Lambda\neq 0$ is a consequence of dS symmetry on quantum level: since dS symmetry is more general than Poincare one then on classical level $\Lambda$ {\it must} be nonzero.
This has nothing to do with gravity, existence or nonexistence of dark energy and with the problem whether
or not  empty space-time should be necessarily flat. 
The parameter $R$
is fully analogous to $c$ and $\hbar$. This parameter {\it should not} be used in pure dS theory and its
only purpose is to get a less general theory (Poincare one) as a formal limit $R\to\infty$ of dS theory.
The question of why $R$ is as it is, is not a matter of fundamental physics since the
answer is: because we want to measure distances in meters. In
particular, there is no guaranty that the CC is really a constant, i.e. does not change with
time. 

At the same time, our derivation depends on two assumptions. The assumption that ${\bf r}=i\partial/\partial {\bf q}$
is the position operator implies that the momentum and position operators are related to each other by the
Fourier transform. As noted in Sec. \ref{intro}, such a choice of the position operator is problematic in
view of the WPS effect. It has been noted that for macroscopic bodies this effect is negligible. However,
from the first principles of quantum theory it is not clear whether there exists a universal choice of the position operator and
whether this operator is needed at all. Another problem is that, since the meaning of time on quantum
level is not clear, the physical meaning of standard prescription on how classical equations of motion arise from quantum 
theory is not clear as well. 

\section{Construction of IRs in discrete basis}
\label{discretebasis}

As noted in Secs. \ref{mainstatement} and \ref{FQT}, for IRs of the so(1,4) algebra it is possible to find a basis such that
all representation operators have only discrete spectrum. This is important for understanding the
relation between standard theory and FQT. By analogy with the method of little group for constructing standard IRs, we first define the rest states
and then the other states can be obtained from them by the action of representation operators.

In spinless case the space of rest states is one-dimensional and its basis consists of only one vector
which we denote as $e_0$. Since ${\bf B}$ is the dS analog of ${\bf P}$ (see Sec. \ref{2particles})
and in the spinless case the angular momentum of the rest state is zero, we define $e_0$ as the
vector satisfying the conditions
\begin{equation}
{\bf B}e_0={\bf J}e_0=0, \quad I_2e_0=(w+9)e_0
\label{e0}
\end{equation}
i.e. $w$ has the meaning of $m_{dS}^2$.

We define $e_1=2{\cal E}e_0$ and 
\begin{equation}
e_{n+1}=2{\cal E}e_n-[w+(2n+1)^2]e_{n-1}
\label{en}
\end{equation}
These definitions make it possible to find $e_n$ for any $n=0,1,2...$.  

We use the notation $J_x=J^1$, $J_y=J^2$,
$J_z=J^3$ and analogously for the operators ${\bf N}$ and ${\bf B}$. Instead of the $(xy)$ components
of the vectors it is convenient to use the $\pm$ components such that $J_x=J_++J_-$,
$J_y=-i(J_+-J_-)$ and analogously for the operators ${\bf N}$ and ${\bf B}$. We now define the elements
$e_{nkl}$ as
\begin{equation}
e_{nkl}=\frac{(2k+1)!!}{k!l!}(J_-)^l(B_+)^ke_n
\label{enkl}
\end{equation}

It can be shown \cite{monograph} that $e_{nkl}$ is the eigenvector of the operator ${\bf B}^2$
with the eigenvalue $4n(n+2)-4k(k+1)$, the eigenvector of the operator ${\bf J}^2$
with the eigenvalue $4k(k+1)$ and the eigenvector of the operator $J_z$ with the eigenvalue $2(k-l)$.
Therefore different vectors $e_{nkl}$ are mutually orthogonal. It can be shown \cite{monograph} that 
the scalar product compatible with 
the Hermiticity of the operators 
$({\cal E},{\bf B},{\bf N},{\bf J})$ can be defined such that 
\begin{equation}
(e_{nkl},e_{nkl})=(2k+1)!C_{2k}^lC_n^kC_{n+k+1}^k\prod_{j=1}^n [w+(2j+1)^2]
\label{norm}
\end{equation}
where $C_n^k=n!/[(n-k)!k!]$ is the binomial coefficient. 
At this point we do not normalize basis vectors to one since, as will be discussed below, the
normalization (\ref{norm}) has its own advantages. At a fixed value of $n$, $k$ takes the values
$k=0,1,...n$, $l$ takes the values $l=0,1,...2k$ and if $l$ and $k$ are not in
this range then $e_{nkl}=0$.

Instead of $l$ we define a new quantum number $\mu =k-l$ which can take values $-k,-k+1,...k$. 
Each element of the representation space can be written as 
$x=\sum_{nk\mu}c(n,k,\mu)e_{nk\mu}$ 
where the set of
the coefficients $c(n,k,\mu)$ can be called the WF in the $(nk\mu)$ representation. 
Assuming that we work with a finite field, a direct calculation (see Ref. \cite{monograph}) shows that
\begin{eqnarray}
&&{\cal E}c(n,k,\mu)=\frac{n-k}{2n}c(n-1,k,\mu)+\frac{n+2+k}{2(n+2)}[w+(2n+3)^2]c(n+1,k,\mu)\nonumber\\
&&J_zc(n,k,\mu)=2\mu c(n,k,\mu)
\label{BJc}
\end{eqnarray}
and for the expressions for other representation operators see Ref. \cite{monograph}.
It is seen from the second expression that the meaning of the quantum number 
$\mu$ is such that $c(n,k,\mu)$
is the eigenfunction of the operator $J_z$ with the eigenvalue $2\mu$, i.e. $\mu$ is standard 
magnetic quantum number.

We use ${\tilde c}(n,k,\mu)$ 
to denote the WF in the basis where the basis elements are normalized to one. Then a direct calculation \cite{monograph}
shows that the action of the representation operators is given by
\begin{eqnarray}
&&{\cal E}{\tilde c}(n,k,\mu)=\frac{1}{2}[\frac{(n-k)(n+k+1)}{n(n+1)}(w+(2n+1)^2)]^{1/2}{\tilde c}(n-1,k,\mu)+\nonumber\\
&&\frac{1}{2}[\frac{(n+1-k)(n+k+2)}{(n+1)(n+2)}(w+(2n+3)^2)]^{1/2}{\tilde c}(n+1,k,\mu)]\nonumber\\
&&N_z{\tilde c}(n,k,\mu)=-\frac{i}{2}[\frac{(k-\mu)(k+\mu)}{(2k-1)(2k+1)(n+1)}]^{1/2}\nonumber\\
&&\{[\frac{(n+k)(n+k+1)}{n}(w+(2n+1)^2)]^{1/2}{\tilde c}(n-1,k-1,\mu)-\nonumber\\
&&[\frac{(n+2-k)(n+1-k)}{n+2}(w+(2n+3)^2)]^{1/2}{\tilde c}(n+1,k-1,\mu)\}-\nonumber\\
&&\frac{i}{2}[\frac{(k+1-\mu)(k+1+\mu)}{(2k+1)(2k+3)(n+1)}]^{1/2}\nonumber\\
&&\{[\frac{(n-k)(n-k-1)}{n}(w+(2n+1)^2)]^{1/2}{\tilde c}(n-1,k+1,\mu)-\nonumber\\
&&[\frac{(n+k+2)(n+k+3)}{n+2}(w+(2n+3)^2)]^{1/2}{\tilde c}(n+1,k+1,\mu)]\}\nonumber\\
&&B_z{\tilde c}(n,k,\mu)=-2[\frac{(k-\mu)(k+\mu)(n+1-k)(n+1+k)}{(2k-1)(2k+1)}]^{1/2}{\tilde c}(n,k-1,\mu)\nonumber\\
&&-2[\frac{(k+1-\mu)(k+1+\mu)(n-k)(n+k+2)}{(2k+1)(2k+3)}]^{1/2}{\tilde c}(n,k+1,\mu)
\label{BJtildec}
\end{eqnarray}
and for the expressions for other representation operators see Ref. \cite{monograph}.

As noted in Secs. \ref{2particles}, the operator ${\bf B}$ is the dS 
analog of the usual momentum ${\bf P}$
such that in Poincare limit ${\bf B}=2R{\bf P}$. At the same time, the operator ${\bf J}$ has the same meaning as in Poincare invariant theory. 

The above expressions are valid in both, standard theory and FQT. One of the difference between those cases is that,
as explained in Sec. \ref{FQT}, in standard theory states are elements of a Hilbert space while in FQT states
are elements of a space over a finite ring or field. In the case of FQT the above expressions do not contain the
characteristic $p$ explicitly. Therefore, as noted in Sec. \ref{FQT}, the correspondence between standard theory
and FQT takes place for states with quantum numbers $(nk\mu)$ such that the absolute value of all those numbers
are much less than $p$.

Another difference between standard theory and FQT follows. In the first case the IR is infinite-dimensional and $n$ can
take the values $0,1,2,...\infty$. On the other hand, as noted in Sec. \ref{FQT}, all IRs in FQT are necessarily finite
dimensional according to the Zassenhaus theorem \cite{Zassenhaus}. It can be shown \cite{monograph} that in the
given case $n$ can take only values $0,1,2,...n_{max}$ where $n_{max}$ is defined by the condition that 
$w+(2n_{max}+3)^2=0$ in $R_p$, i.e. $w+(2n_{max}+3)^2$ is a multiple of p. We will see below that $p$ is a huge
number and therefore for real bodies $n_{max}$ is huge too. Therefore if the WF is not zero only for the values
of $n$ such that $n\ll n_{max}$ then, in the case of IRs, the supports of the $n$-distribution in standard theory 
and FQT are the same. Therefore the supports of the $k$- and $\mu$-distributions are the same too.

Consider now the semiclassical approximation in the normalized basis of standard theory. In view of the
usual understanding of the structure of semiclassical WFs (see e.g. Ref. \cite{monograph})
a necessary condition for the semiclassical approximation is that
the quantum numbers $(nk\mu)$ are much greater than 1. We assume that a state is semiclassical 
if its WF has the form
\begin{equation}
{\tilde c}(n,k,\mu)=a(n,k,\mu)exp[i(-n\varphi+k\alpha -\mu\beta)]
\label{qclwf}
\end{equation}
where $a(n,k,\mu)$ is the amplitude, which is not small only in some vicinities of $n=n_0$, $k=k_0$ and 
$\mu=\mu_0$.
We also assume that when the quantum numbers $(nk\mu)$ change by one, the main contribution
comes from the rapidly oscillating exponent. Then, as follows from the first expression 
in Eq. (\ref{BJtildec}), the action of the dS energy operator can be written as
\begin{eqnarray}
{\cal E}{\tilde c}(n,k,\mu)\approx\frac{1}{n_0}[(n_0-k_0)(n_0+k_0)(w+4n_0^2)]^{1/2}cos\varphi
{\tilde c}(n,k,\mu)
\label{calE}
\end{eqnarray}
Therefore the semiclassical WF is approximately the eigenfunction of the dS
energy operator with the eigenvalue
$$\frac{1}{n_0}[(n_0-k_0)(n_0+k_0)(w+4n_0^2)]^{1/2}cos\varphi.$$
When $n_0\gg k_0$ and $\varphi\ll 1$ the eigenvalue equals $(w+4n_0^2)^{1/2}$. Since
$w=m_{dS}^2$, this result shows that $n$ is the dS analog of the absolute value of the momentum,
i.e. in Poincare approximation $n\approx R|{\bf P}|$.

\section{System of two particles in discrete basis}
\label{2particlesdiscrete}

Consider now a system of two free particles in dS theory. Then the two-particle operator $M_{ab}$ is a sum
of the corresponding single-particle operators and the two-body Casimir operator can be defined by Eq. (\ref{casimir})
with the two-particle operators $M_{ab}$. 
By analogy with the single-particle case, one can define the two-body operator $W$ which is an analog 
of the quantity $w$:
\begin{equation}
I_2=W-{\bf S}^2+9
\label{W}
\end{equation}
where ${\bf S}$ is the two-body spin operator. 

By analogy with standard theory, it is convenient to consider the two-body mass operator if individual particle
dS momenta $n_1$ and $n_2$ are expressed in terms of the total and relative dS momenta $N$ and $n$. In the c.m. frame
we can assume that ${\bf B}_1$ is directed along the positive direction of the $z$ axis and then ${\bf B}_2$ is
directed along the negative direction of the $z$ axis. Therefore the quantum number $N$ characterizing the total
dS momentum can be defined as $N=n_1-n_2$. In nonrelativistic theory the relative
momentum is defined as ${\bf q}=(m_2{\bf p}_1-m_1{\bf p}_2)/(m_1+m_2)$ and in relativistic theory as 
${\bf q}=(E_2{\bf p}_1-E_1{\bf p}_2)/(E_1+E_2)$.
Therefore, taking into account the fact that in the c.m. frame the particle momenta are directed in opposite
directions, one might define $n$ as $n=(m_2n_1+m_1n_2)/(m_1+m_2)$ or $n=(E_2n_1+E_1n_2)/(E_1+E_2)$. These
definitions involve Poincare masses and energies. Another possibility is $n=(n_1+n_2)/2$. In all these
cases we have that $n\to (n+1)$ when $n_1\to (n_1+1),\,\,n_2\to (n_2+1)$ and 
$n\to (n-1)$ when $n_1\to (n_1-1),\,\,n_2\to (n_2-1)$. In what follows, only this feature is important. 

Consider the space of functions ${\tilde c}(n)$ such that 
$$\sum_{n=0}^{\infty}|{\tilde c}(n)|^2<\infty$$
Let ${\cal B}$ be the operator which acts in this space as 
\begin{equation}
{\cal B}{\tilde c}(n)=\frac{1}{2}[{\tilde c}(n+1)+{\tilde c}(n-1)]
\label{calB}
\end{equation}
and $G=1-{\cal B}$. As shown in Ref. \cite{monograph}, in the approximation 
when $n_j\gg k_j$ ($j=1,2$)
\begin{equation}
W=W_0-2(w_1+4n_1^2)^{1/2}(w_2+4n_2^2)^{1/2}G
\label{WW}
\end{equation}
where 
\begin{equation}
W_0=w_1+w_2+2(w_1+4n_1^2)^{1/2}(w_2+4n_2^2)^{1/2}-2{\bf B}_1{\bf B}_2
\label{W0}
\end{equation}
This operator can be represented as $W_0=4R^2M_0^2$ where $M_0^2=(p_1+p_2)^2$,
$p_j$ ($j=1,2$) is standard Poincare four-momentum of particle $j$ and therefore $M_0^2$
is the free mass operator squared in Poincare invariant theory.

Since classical mechanics works with a very high accuracy at macroscopic level, one might think that
the validity of semiclassical approximation at this level is beyond any doubts. However, this problem has not been investigated quantitatively. In quantum theory a
physical quantity is treated as semiclassical if its  uncertainty is much less than its
mean value. Consider WFs describing the motion of macroscopic bodies as
a whole (say the WFs of the Sun, the Earth, the Moon etc.). It is obvious that uncertainties of
coordinates in these WFs are much less than the corresponding macroscopic dimensions. 
What are those uncertainties for the Sun, the Earth, the 
Moon, etc.? What are the uncertainties of their momenta?

If $A$ is a physical quantity then we use $\Delta A$ to denote the uncertainty of this quantity in some state. 
In standard quantum mechanics, the validity of  
semiclassical approximation is defined by the product $\Delta r \Delta p$ while each uncertainty by
itself can be rather large. It is known that if, for example, the coordinate and momentum WFs are Gaussian
then $\Delta r \Delta p$ is of the order of unity. On the other hand, as noted in Sec. \ref{intro}, the validity of  
standard position operator is problematic. Do we know what scenario for the distribution of momenta and coordinates 
takes place for macroscopic bodies?

In view of the correspondence between standard theory and FQT we will consider only WFs with a
finite support.  One might think that a necessary condition for the validity of
semiclassical approximation is that the exponent in the semiclassical WF makes many oscillations
in the region where the WF is not small. We will consider WFs $\psi(n)$ 
containing $exp(-i\varphi n)$ such that $\psi(n)$ can be different from zero only if $n\in [n_{min},n_{max}]$.
Then, if $\delta=n_{max}-n_{min}$, the exponent makes $|\varphi| \delta/2\pi$ oscillations on $[n_{min},n_{max}]$ and $\varphi$ should satisfy the condition $|\varphi|\gg 1/ \delta$. 
The problem arises whether this condition is sufficient. 

As already noted, a quantity can be semiclassical only if its mean value is much greater than its
uncertainty. In particular, a quantity cannot be semiclassical if its mean value is zero or very small.
In Poincare theory the exponent is written as $exp(-ipr)$. Since $n$ is the Poincare analog of $Rp$, one
might think that $\varphi$ is the Poincare analog of $r/R$. Since Poincare limit is treated as $R\to\infty$,
in Poincare limit $\varphi$ is not zero only for cosmological distances. This poses the
problem whether $\varphi$ can be semiclassical for non-cosmological distances. The dS analog of $\Delta r \Delta p$
is $\Delta\varphi \Delta n=\Delta\varphi \delta$ and the problem arises whether there exist states where this
product is of the order of unity.

In Ref. \cite{monograph} we discussed in detail the choice of the two-body relative distance operator. Since
the functions ${\tilde c}(n)$ are discrete and have a finite support, we now do not have an option to choose
the momentum and coordinate WFs Gaussian. As shown in Ref. \cite{monograph}, if the coordinate $r$ is treated as
$\varphi R$ and $r\ll R$ then even for favorable scenarios $\Delta\varphi$ is of the order of $1/\delta^{1/2}$.
Therefore $\Delta\varphi \delta$ is a very large value of the order of $\delta^{1/2}$ and this is unacceptable.
We argue that the coordinate is semiclassical if $exp(-i\varphi n)$ is replaced by
$exp(-i\theta n)$ where $\theta=const/(\delta\varphi)^{1/2}$. 

The mean value of the operator $W$ can be written as $\overline{W}=4R^2M_0^2+{\overline {\Delta W}}$ where the
last term is the dS correction to the result in Poincare theory. 
If the exponent in the internal WF is $exp(-i\varphi n)$, $\varphi$
is understood as $r/R$ and $\varphi\ll 1$ then as follows from Eq. (\ref{WW}) \cite{monograph}
\begin{equation}
{\overline {\Delta W}}=-4R^2[(m_1^2+{\bf p}_1^2)(m_2^2+{\bf p}_2^2)]^{1/2}\varphi^2
\label{CCdiscrete}
\end{equation}
As noted above, this can be justified if $r$ is cosmological but still much less than the parameter $R$.
As follows from Eq. (\ref{CCdiscrete}), in the nonrelativistic approximation we get the same result as in Eq. (\ref{2M}).

If the exponent in the internal WF is $exp(-i\theta n)$, $\theta\ll 1$ then as follows from Eq. (\ref{WW}) \cite{monograph}
\begin{equation}
{\overline {\Delta W}}=-const^2[(w_1+4n_1^2)(w_2+4n_2^2)]^{1/2}\frac{\delta_1+\delta_2}
{\delta_1\delta_2|\varphi|}
\label{preNewton}
\end{equation}
and the result for the classical nonrelativistic Hamiltonian is
\begin{equation}
H({\bf r}, {\bf q}) =\frac{{\bf q}^2}{2m_{12}} - \frac{m_1m_2Rconst^2}{2(m_1+m_2)r}(\frac{1}{\delta_1}+\frac{1}{\delta_2})
\label{preNewton2}
\end{equation}
where $\delta_j$ ($j=1,2$) is the width of the $n$-distribution for particle $j$.
We see that the correction disappears if the width of the 
dS momentum distribution for each body becomes very large. In standard theory (over complex numbers)
the only limitation is that the width of the dS momentum distribution should be much less
than the mean value of this momentum. This is not a serious restriction and the width can be arbitrarily large.
In the next section we argue that in FQT it is natural that
the width of the momentum distribution for a macroscopic body is inversely proportional to its mass.
Then we recover the Newton gravitational law. Namely, if
\begin{equation}
\delta_j=\frac{R}{m_jG'}\quad (j=1,2), \quad const^2G'=2G 
\label{deltaj}
\end{equation}
then
\begin{equation}
H({\bf r}, {\bf q}) =\frac{{\bf q}^2}{2m_{12}} - G\frac{m_1m_2}{r}
\label{Newton}
\end{equation}
In Ref. \cite{monograph} we also discussed relativistic corrections to the Newton law.
We conclude that in our approach gravity is not an interaction but simply the dS correction to standard free nonrelativistic Hamiltonian. 

As noted above, classical equations of motions can be obtained from the Hamiltonian in different ways. 
If $m_2\gg m_1$ then the Newton law for particle 1 can be obtained from the single-particle operators 
discussed in the preceding section. In this case $\delta_1\gg \delta_2$ and, as follows from Eqs. (\ref{preNewton2}) and (\ref{deltaj}),
\begin{equation}
H({\bf r}, {\bf q}) =\frac{{\bf q}^2}{2m_1} - G\frac{m_1m_2}{r}
\label{1bodyNewton}
\end{equation}
if
\begin{equation}
\delta_2=\frac{Rconst^2}{2m_2G}
\label{deltaM}
\end{equation}
Therefore for particle 1 the presence of the heavy body is manifested such that the single-particle width $\delta_1$ should be
replaced by the width of the $n$-distribution which equals $\delta_2$.

\section{Semiclassical states in FQT}
\label{S9}

For any new theory there should exist a correspondence principle that at some conditions 
this theory and standard well tested one should give close predictions. As noted in Sec. \ref{FQT}, we treat standard quantum theory
as a special case of FQT in the formal limit $p\to\infty$. A detailed discussion of FQT has been given
in Refs. \cite{lev4,monograph,lev2}. Here we describe only basic facts needed for further presentation.  

As noted in Sec. \ref{discretebasis}, a single-particle WF can be written as 
$$x=\sum_{nk\mu}c(n,k,\mu)e_{nk\mu}$$
For the validity of semiclassical approximation the condition 
\begin{equation}
\sum_{nk\mu} ||e(n,k,\mu)||^2|c(n,k,\mu)|^2\ll p
\label{llp}
\end{equation}
should be satisfied. A detailed analysis in Ref. \cite{monograph} shows that if $n\gg k$ then this
condition can be satisfied if 
\begin{equation}
\delta lnw \ll lnp
\label{lnrho}
\end{equation}
Therefore not only the number $p$ should be very large, but even $lnp$ should be very large. 
Note that in finite mathematics there is no logarithm but in number theory it is rather often used
for estimations. For example, the famous prime number theorem describing the asymptotic distribution of primes
 involves logarithm.

For elementary particles the condition (\ref{lnrho}) is always valid without any doubts. Consider now what happens in
the case of a macroscopic body which consists of many elementary particles. 
In quantum theory, state vectors of a system of N bodies belong to the Hilbert space which is the 
tensor product of single-body Hilbert spaces. This means that state vectors of the $N$-body systems
are all possible linear combinations of functions
\begin{equation}
\psi(n_1,k_1,l_1,...n_N,k_N,l_N)=\psi_1(n_1,k_1,l_1)\cdots \psi_N(n_N,k_N,l_N)
\label{tprod}
\end{equation}
By definition, the bodies do not interact if all representation operators of the symmetry algebra for
the $N$-body systems are sums of the corresponding single-body operators. For example, the energy operator
${\cal E}$ for the $N$-body system is a sum ${\cal E}_1+{\cal E}_2+...+{\cal E}_N$ where the operator
${\cal E}_i$ ($i=1,2,...N$) acts non-trivially over its "own" variables $(n_i,k_i,l_i)$ while over other
variables it acts as the identity operator.  

If we have a system of noninteracting bodies in standard quantum theory, each $\psi_i(n_i,k_i,l_i)$ in
Eq. (\ref{tprod}) is fully independent of states of other bodies. However, in FQT the
situation is different. Here, as noted above,  a necessary condition for the WF to have a probabilistic 
interpretation is given by Eq. (\ref{lnrho}), and for elementary particles this is not a serious restriction. 
However, if a
system consists of $N$ components, a necessary condition that the WF of the system has a
probabilistic interpretation is
\begin{equation}
\sum_{i=1}^N \delta_i lnw_i \ll lnp
\label{70}
\end{equation} 
where $\delta_i=\Delta n_i$ and $w_i=4R^2m_i^2$ where $m_i$ is the mass of the subsystem $i$. 
This condition shows that in FQT the greater the number of components is,
the stronger is the restriction on the width of the dS momentum distribution for each component.
This  is a crucial difference between standard theory and FQT. A naive explanation is that if $p$ is
finite, the same set of numbers which was used for describing one body is now shared between $N$ bodies.
In other words, if in standard theory each body in the free $N$-body system does not feel the presence of
other bodies, in FQT this is not the case. This might be treated as an effective interaction in the free
$N$-body system.

The existing quantum theory does not make it possible to reliably calculate the width of the total dS momentum
distribution for a macroscopic body and at best only a qualitative estimation of this quantity can be
given. The above discussion shows that the greater the mass of the macroscopic body is, the stronger is
the restriction on the dS momentum distribution for each subsystem of this body. Suppose that a
body with the mass $M$ can be treated as a composite system consisting of similar subsystems with the mass $m$.
Then the number of subsystems is $N=M/m$ and, as follows from Eq. (\ref{70}), the width $\delta$ of their dS momentum
distributions should satisfy the condition $N\delta lnw\ll lnp$ where $w=4R^2m^2$. Since the greater the value 
of $\delta$ is, the more accurate is the
semiclassical approximation, a reasonable scenario is that each subsystem tends to have the maximum possible
$\delta$ but the above restriction allows to have only such value of $\delta$ that it is of the order 
of magnitude not exceeding $lnp/(Nlnw)$. 

The next question is how to estimate the width of the total dS momentum distribution for a macroscopic body.
For solving this problem one has to change variables from individual dS momenta of subsystems to total and
relative dS momenta. Now the total dS momentum and relative dS momenta will have their own momentum distributions
which are subject to a restriction similar to that given by Eq. (\ref{70}). If we assume that all the variables
share this restriction equally then the width of the total momentum distribution also will be a quantity not
exceeding $lnp/(Nlnw)$. Suppose that $m=N_1m_0$ where $m_0$ is the nucleon mass. The value of $N_1$ should be 
such that our subsystem still can be described by semiclassical approximation. Then the estimation of $\delta$ is
\begin{equation}
\delta=N_1m_0lnp/[2Mln(2RN_1m_0)]
\label{estimation}
\end{equation}
Note that if standard masses in this expression are replaced by the corresponding dS masses then the expression 
will not contain $R$.
Suppose that $N_1$ can be taken to be the same for all macroscopic bodies.
For example, it is reasonable to expect that when $N_1$ is of the order of $10^3$, the subsystems still can be
described by semiclassical approximation but probably this is the case even for smaller values of $N_1$.

In summary, although calculation of the width of the total dS momentum distribution for a macroscopic body is
a very difficult problem, FQT gives a reasonable qualitative explanation why this quantity is inversely
proportional to the mass of the body. With the estimation (\ref{estimation}), the result given by 
Eq. (\ref{preNewton2}) can be written in the form (\ref{Newton}) where 
\begin{equation}
G=\frac{2const\, Rln(2RN_1m_0)}{N_1m_0lnp}
\label{G}
\end{equation} 

In Sec. \ref{symmetry} we argued that in theories based on dS invariance there should be no dimenful quantities.
In particular, neither the gravitational nor cosmological constant can be fundamental. In units $\hbar/2=c=1$, 
the dimension of $G$ is $length^2$ and its
numerical value is $l_P^2$ where $l_P$ is the Planck length ($l_P\approx 10^{-35}m$). Equation (\ref{G})
is an additional indication that this is the case since $G$ depends on $R$ (or the cosmological constant) and
there is no reason to think that it does not change with time. Since $G_{dS}=G\Lambda$ is dimensionless
in units $\hbar/2=c=1$, this quantity should be treated as the gravitational constant in dS theory. 
If $\mu=2Rm_0$ is the dS nucleon mass then Eq. (\ref{G}) can be written as
\begin{equation}
G_{dS}=\frac{12const\, ln(N_1\mu )}{N_1\mu lnp}
\label{GLambda}
\end{equation} 
As noted in Sect. \ref{2particles}, standard cosmological constant problem arises when one tries to explain the value of
$\Lambda$ from quantum theory of gravity assuming that this theory is QFT, $G$ is fundamental and dS symmetry is a manifestation of dark energy (or other fields) on flat Minkowski background. Such a theory contains strong divergences and the result depends on the
value of the cutoff momentum. With a reasonable assumption about this value, the quantity $\Lambda$ is of the order of
$1/G$ and this is reasonable since $G$ is the only parameter in this theory. Then, as noted above, $\Lambda$ is by 122
 orders of
magnitude greater than its experimental value. However, in our approach we have an
additional fundamental parameter $p$. Equation (\ref{GLambda}) shows that $G\Lambda$ is
not of the order of unity but is very small since not only $p$ but even $ln p$ is very large. For a rough estimation, we assume that 
the values of $const$ and $N_1$ in this expression are of the order of unity. Then if, for example, $R$ is of the order 
of $10^{26}meters$,
we have that $\mu$ is of the order of $10^{42}$ and $lnp$ is of the order of $10^{80}$. Therefore $p$ is a huge number of
the order of $exp(10^{80})$. As noted in Sec. \ref{2particles}, the value of $R$ may be even much greater than $10^{26}meters$ and
in that case the value of $p$ will be even much greater than $exp(10^{80})$. 

Concluding this section we would like to make remarks about the hierarchy of physical theories. As discussed in Secs. 
\ref{fundamentaltheories}, \ref{mainstatement}, \ref{symmetry} and \ref{2particles},
transition from a more general theory to a less general one can be accomplished such that the more general theory can be written with some
finite parameter and the less general theory is obtained as a formal limit when this parameter goes to zero or infinity. From this point
of view, FQT is the most general theory since all other theories can be obtained from FQT by this procedure. Since FQT is based on finite
mathematics it should depend on a finite parameter $p$ which is roughly the greatest possible number in the theory and no physical quantity
can exceed this number. As noted above, in our approach gravity is a consequence of the fact that $p$ is finite. 
It is also obvious that FQT should not depend on any dimensionful parameters. When we take a formal limit $p\to\infty$ we
obtain standard dS or AdS theories. They still do not depend on dimensionful quantities. However, when we introduce the quantity $R$ and take the
limit $R\to\infty$ we obtain quantum Poincare theory in which the dimensionful parameters can have only the dimension of $length$ or its powers.
When we take the limit $\hbar\to 0$ or $c\to\infty$ we obtain less general theories with greater number of dimensions. The less general theory
is classical nonrelativistic theory which depend on dimensions $(kg,m,s)$.

\section{Classical equations of motions in FQT}
\label{classicalFQT}

\subsection{One-dimensional model}
\label{1dim}

Consider a system of two particles with the masses $m_1$ and $m_2$ such that $m_2\gg m_1$. Then, as noted at the end of Sec. \ref{2particlesdiscrete},
particle 1 can be considered in the framework of single-particle problem but the width of the $n_1$ distribution should be
replaced by the width of the $n$ distribution which equals $\delta=\delta_2$. For simplicity we will consider the case when on classical
level the particle is moving along the $z$-axes. The corresponding semiclassical WF is the eigenstate of the operator $J_z$ with the eigenvalue 
$\mu=0$ and such that the parameter $\alpha$ in Eq. (\ref{qclwf}) is zero or $\pi$. 
Our goal is to obtain classical results without using standard semiclassical approximation, position operators and time but proceeding only
from quantum states. However, the semiclassical results give a hint that if $k\ll n$ then a simple case which we can consider is the one-dimensional model
where the WF $c(n)$ depends only on $n$ and, as follows from the first expression in Eq. (\ref{BJc}) 
\begin{eqnarray}
{\cal E}c(n)=\frac{1}{2}c(n-1)+\frac{1}{2}[w+(2n+3)^2]c(n+1)
\label{1E}
\end{eqnarray}
Although we work in FQT, it will be helpful to compare the results with those obtained in standard theory because our physical
intuition is based on that theory. Here, as follows from
Eq. (\ref{BJtildec}), the dS energy operator acts on the normalized WF as
\begin{equation}
{\cal E}{\tilde c}(n)=\frac{1}{2}[(w+(2n+1)^2)]^{1/2}{\tilde c}(n-1)+
\frac{1}{2}[w+(2n+3)^2)]^{1/2}{\tilde c}(n+1)
\label{1Etilde}
\end{equation}

For the correspondence with standard theory, in FQT it is desirable to work with least possible
numbers in order to avoid comparisons modulo $p$ whenever possible. We now use $n_1$ and $n_2$
to define the minimum and maximum values of $n$ in the support of $c(n)$. Then by using the
fact that the space of states is projective, as follows from Eq. (\ref{norm}), the normalization of the
elements $e_n$ can be chosen as
\begin{equation}
(e_n,e_n)=\prod_{j=n_1+1}^n [w+(2j+1)^2]\quad (n\in[n_1, n_2])
\label{1norm}
\end{equation}
Then up to a normalization factor  the relation between the WFs in FQT and in standard theory can be written in the form
\begin{equation}
{\tilde c}(n_2-l)=c(n_2-l)\{\prod_{m=0}^{l-1}[w+(2n_2-2m+1)^2]\}^{-1/2}
\label{ctildec}
\end{equation}
where $l=n_2-n$.

Since $c(n)$ has a finite support it cannot be the eigenstate of the operator ${\cal E}$. For example,
$c(n_2+1)=0$ but, as follows from Eq. (\ref{1E}), ${\cal E}c(n_2+1)=c(n_2)/2\neq 0$. 
Analogously $c(n_1-1)=0$ but, as follows from Eq. (\ref{1E}), ${\cal E}c(n_1-1)=[w+(2n_1+1)^2]c(n_1)/2\neq 0$. 
We will see below
that the uncertainty of ${\cal E}$ is minimal when ${\cal E}c(n)=\lambda c(n)$ for $n\in [n_1,n_2]$.
This condition can be satisfied if the expression describing $c(n)$ at $n\in [n_1,n_2]$ is such that $c(n_1-1)=0$ and $c(n_2+1)=0$. 

Since the norm of $e_n$ is maximal when $n=n_2$, we want to work with least possible numbers, 
the states are projective, the minimum possible value of $c(n_2)$ in FQT is $c(n_2)=\pm 1$ then we
choose $c(n_2)=1$. Then, as follows from Eq. (\ref{1E}), for $n\in [n_1,n_2]$ all the values $c(n)$ can be found consecutively:
\begin{equation}
c(n-1)=2\lambda c(n) - [w+(2n+3)^2]c(n+1)
\label{c(n)}
\end{equation}
In particular, $c(n_2-1)=2\lambda$, $c(n_2-2)=4\lambda^2-W$ etc. However, it is
problematic to find an explicit expression for $c(n)$ if $n$ is arbitrary.

In the nonrelativistic case $w\gg n_2^2$ and for semiclassical WFs $\delta=(n_2-n_1)\ll n_2$. So one might think that a good approximation is to neglect the variations of $[w+(2n+1)^2]$ at $n\in [n_1,n_2]$ and
consider the following approximation of Eq. (\ref{c(n)}):
\begin{equation}
c(n-1)=2\lambda c(n) -Wc(n+1)
\label{cW}
\end{equation} 
 where $W=w+(2n_2+1)^2$. Then it is easy to prove by induction that
\begin{equation}
c(n_2-l)=\sum_{m=0}\frac{(-1)^m (l-m)!}{m!(l-2m)!}(2\lambda)^{l-2m}W^m
\label{cl}
\end{equation}
where the upper limit is defined by the condition that 
$1/(l-2m)!=0$ if $l<2m$. 
As follows from Eq. (\ref{ctildec}), in this approximation
\begin{equation}
{\tilde c}(n_2-l)=C(l)=C(l,x)=\sum_{m=0}\frac{(-1)^m (l-m)!}{m!(l-2m)!}(2x)^{l-2m}
\label{cl}
\end{equation}
where $x=\lambda/W^{1/2}$. This is the Gegenbauer polynomial which in the literature is
denoted as $C_l^1(x)$, and it is known that if $x=cos\theta$ then $C(l)=sin((l+1)\theta)/sin\theta$.
Since the notation $C_n^k$ is also used for binomial coefficients we will use for the
Gegenbauer polynomial $C_n^k(x)$ the notation $G_n^k(x)$.

Suppose that $sin((\delta +2)\theta)=0$. Then $(\delta+2)\theta=k\pi$ where $k$ is an integer,  
$sin((\delta+1)\theta )=(-1)^{k+1}sin\theta$ and 
\begin{equation}
Norm^2=\sum_{l=0}^{\delta}C(l)^2=\frac{1}{sin^2\theta}\sum_{l=0}^{\delta}sin^2((l+1)\theta)=
\frac{\delta+2}{2sin^2\theta}
\label{Norm}
\end{equation}
In this case ${\cal E}{\tilde c}(n)=\lambda {\tilde c}(n)$ for all 
$n\in [n_1,n_2]$, $\lambda$ is exactly the mean value of the operator ${\cal E}$: 
\begin{equation}
\bar{{\cal E}}=\frac{1}{Norm^2}({\tilde c},{\cal E}{\tilde c})=\frac{1}{Norm^2}\sum_{n=n_1}^{n_2} 
{\tilde c}(n){\cal E}{\tilde c}(n)=\lambda ,
\label{barE}
\end{equation}
and the uncertainty of ${\cal E}$ is
\begin{equation}
\Delta {\cal E}=\frac{1}{Norm}({\tilde c},({\cal E}-\bar{{\cal E}})^2{\tilde c})^{1/2}=
\frac{1}{Norm}||({\cal E}-\bar{{\cal E}}){\tilde c}||=(\frac{W}{\delta+2})^{1/2}|sin\theta|
\label{DeltaE}
\end{equation}

As follows from Eq. (\ref{BJc}), if $k\ll n$ then the dS energy of the particle which is far from other
particles approximately equals ${\cal E}\approx \pm W^{1/2}$ and, as follows from Eqs. (\ref{HP})
and (\ref{1bodyNewton}), for nonrelativistic particles the effective interaction gives a small correction to ${\cal E}$.
Therefore $\lambda /W^{1/2}$ is close to 1 but is less than 1.
Hence one can choose $\theta$ such that $cos\theta=\lambda /W^{1/2}$, $\theta$ is small and $\theta>0$. 
Then, as follows from Eq. (\ref{DeltaE}), 
$\Delta {\cal E}/{\bar {\cal E}}\approx sin\theta/\delta^{1/2}$
is very small because $\delta$ is very large and $sin\theta$ is small. Indeed, a simple estimation shows
that if the kinetic and potential energies are of the same order then $\theta$ is of the order of $v/c$
and for the cosmological repulsion $\theta$ is of the order of $r/R$. As a consequence, the particle state
is strongly semiclassical.  

Another possible choice of the WF follows. We do not require that the condition $({\cal E}-\lambda)c(n)=0$ should
be satisfied at all $n\in [n_1,n_2]$,
choose an arbitrary value for $c(n_2-1)$ and find the values of $c(n)$ at $n=n_2-2,...,n_1$ from 
Eq. (\ref{cW}). Then in general the condition $({\cal E}-\lambda)c(n)=0$ will be satisfied only for $n\in [n_1+1,n_2-1]$. 
In particular, if $c(n_2-1)=\lambda$ then it follows from from Eqs. (\ref{ctildec}) and (\ref{cW}) that
${\tilde c}(n_2-l)=cos(l\theta)$. In that case the quantity ${\Delta\cal E}/\bar{{\cal E}}$ will be greater than
in the case of Eq. (\ref{DeltaE}) but will also be of the order not greater than $1/\delta^{1/2}$, i.e. very small.
We conclude that the requirement that the dS energy should be strongly semiclassical does not impose strong
restrictions on the WF.

A problem arises whether it is indeed a good approximation to neglect the variations of $[w+(2n+1)^2]$ at $n\in [n_1,n_2]$.
In what follows we describe two attempts to find the exact solution.

Consider this problem in standard theory and define
$$f(l)=\frac{w+(2(n_2-l)+1)^2}{w+(2n_2+1)^2}, \quad F(l)=[\prod_{m=0}^{l-1}f(m)]^{-1}C(l)$$
Then $F(l)=C(l)$ if $l=0,1$, $F(l) \neq C(l)$ at $l\geq 2$ and, as follows from Eq. (\ref{1Etilde})
\begin{equation}
F(l+1)=2cos\theta F(l) -f(l-1)F(l-1)
\label{F(l)}
\end{equation}
It is obvious that $F(l)\approx C(l)$ for $l\ll \delta$ but the problem is whether the approximate equality
takes place if $l$ is of the order of $\delta$.

We define $S(k,l)=\sum f(i_1)...f(i_l)$ where the sum is taken over all products of $l$ multipliers such that
$S(k,0)=S(0,1)=1$, $S(k,l)=0$ if $k>0$ and $k<l$, the indices $i_1,...i_l$ can take the values $0,1,...k$ in the ascending order and the difference between any
value and the previous one is greater or equal 2. 
Then it can be easily proved by induction that 
\begin{equation}
S(l,m)=S(l-1,m)+f(l)S(l-2,m-1)
\label{S(l,m)}
\end{equation}
We consider the first case discussed above, i.e. $c(n_2)=1$ and $c(n_2-1)=2\lambda$.  
Then it can be proved by induction that, as follows from Eq. (\ref{S(l,m)}), the solution of Eq. (\ref{F(l)}) is 
\begin{equation}
F(l)=\sum_{m=0}^{[l/2]}(-1)^m (2x)^{l-2m}S(l-2,m)
\label{F(l)B}
\end{equation}
where $[l/2]$ is the integer part of $l/2$.

Since we assume that $l\ll n_2$ then $f(l)\approx 1-ly$ where $y=4(2n_2+1)/W$. We assume that if
$l$ is of the order of $\delta$ then the approximate expression for $S(l,m)$ is
\begin{equation}
S(l,m)=\sum_{s=0}^m a(l,m,s), \quad a(l,m,s)=\frac{ (-y/2)^s(l+2-m)!l!}{(l+2-2m)!s!(m-s)!(l-s)!}
\label{S(l,mB)}
\end{equation}
It follows from this expression that only the values of $m\leq (l/2+1)$ contribute to the sum and 
$a(l,m,s+1)/a(l,m,s)=-y(m-s)(l-s)/[2(s+1)]$.

The value of $W$ is the Poincare analog of the energy squared: $W=4R^2(m^2+{\bf p}^2)$, $n_2$ is the Poincare analog of
$R|{\bf p}|$ and, as follows from Eq. (\ref{deltaM}), $\delta$ is of the order of $R/r_g$ where $r_g$ is the gravitational (Schwarzschild)
radius of the heavy body. Then if $l$ is of the order of $\delta$ and $R$ is of the order of $10^{26}meters$ then 
$yl^2 \ll 1$. However, as
noted above, the value of $R$ may be much greater than $10^{26}meters$, Poincare limit is defined as $R\to\infty$ and in the formal limit
$R\to\infty$, $y\delta^2\to\infty$. So if $m$ is of the order $l$ and $s\ll m$ then it is possible that 
$a(l,m,s+1)\gg a(l,m,s)$ but if $s$ if of the order of $m$ then $a(l,m,s+1)\ll a(l,m,s)$.

A direct calculation using Eq. (\ref{S(l,mB)}) gives
\begin{eqnarray}
&& S(l-1,m)+f(l)S(l-2,m-1)=\sum_{s=0}^m b(l,m,s), \nonumber\\
&&b(l,m,s)=a(l,m,s)[1+\frac{s(s-1)(l-s)}{l(l-1)(l+2-m)}]
\label{S(l,mC)}
\end{eqnarray}
Therefore Eq. (\ref{S(l,m)}) is satisfied with a high accuracy and Eq. (\ref{S(l,mB)}) is a good approximate
expression for $S(l,m)$.

As follows from Eqs. (\ref{F(l)}) and (\ref{S(l,mB)}), the expression for $F(l)$ can be represented as
\begin{equation}
F(l)=\sum_{s=0}^{[l/2]} (y/2)^s\frac{(l-2)!}{(l-2-s)!}\sum_{m=0}^{[\nu/2]}(-1)^m(2x)^{\nu-2m}\frac{(s+1)_{\nu-m}}{(\nu-2m)!m!} 
\label{F(l)C}
\end{equation} 
where $\nu=l-2s$ and $n_k=n(n-1)...(n-k+1)$ is the Pohhammer symbol. The last sum in this expression is the Gegenbauer polynomial 
$G_{\nu}^{s+1}(x)$ and therefore
\begin{equation}
F(l)=\sum_{s=0}^{[l/2]} (y/2)^s\frac{(l-2)!}{(l-2-s)!}G_{\nu}^{s+1}(x) 
\label{F(l)D}
\end{equation} 
Finally, by using the asymptotic expression for the Gegenbauer polynomial $G_{\nu}^{s+1}(x)$ when $\nu$ is large we
get
\begin{equation}
F(l)=\sum_{s=0}^{[l/2]}(y/4)^s\frac{(l-2)!(l-s)!}{(l-2-s)!s!(l-2s)!}\frac{cos[(l-s+1)\theta -(s+1)\pi/2]}{sin\theta^{s+1}}
\label{F(l)E}
\end{equation} 

If this expression is represented as $F(l)=\sum_s a(l,s)$ then for $l$ of the order of $\delta$ and $s\ll l$,
$a(l,s+1)/a(l,s)$ is of the order of $yl^2/sin\theta$. As noted above, the quantity $yl^2$ can be very large and
therefore the quantity $yl^2/sin\theta$ can be even larger, especially in cases when $\theta$ is of the order of $r/R$.
We see that even for the choice $c(n_2-1)=2\lambda$ understanding qualitative features of the
solution of Eq. (\ref{1E}) is very difficult. In addition, as noted above, the WF is strongly semiclassical for
other choices of $c(n_1-1)$. Therefore it is a great problem to understand what conditions govern the choice
of the semiclassical WF. 

The second attempt to find the exact solution follows. Consider the function 
\begin{equation}
{\tilde c}(n_2-l)=const\cdot  cos(\alpha(l)), \quad \alpha(l)=\sum_{m=1}^l arccos(\frac{\lambda}{[w+(2(n_2-m)+3)^2]^{1/2}})
\label{approxA}
\end{equation} 
where $const$ is a normalizing coefficient. When the variations of 
$[w+(2n+1)^2]$ at $n\in [n_1,n_2]$ are neglected this function becomes ${\tilde c}(n_2-l)=const\cdot cos(l\theta)$, i.e.
the approximate solution discussed above.
As follows from Eqs. (\ref{1Etilde}) and (\ref{approxA})
\begin{eqnarray}
&&{\cal E}{\tilde c}(n_2-l)=\lambda {\tilde c}(n_2-l) +\frac{1}{2}const\cdot sin((\alpha(l))\{[w+(2(n_2-l)+3)^2-\lambda^2]^{1/2}-\nonumber\\
&&[w+(2(n_2-l)+1)^2-\lambda^2]^{1/2}\}
\label{EA}
\end{eqnarray}
The presence of the second term in the r.h.s. shows that the function given by Eq. (\ref{approxA}) is not the exact solution. Typically this term is much less than the first one but this is not the case when $cos(\alpha(l))$ is small.

Analogously the function ${\tilde c}(n_2-l)=const\cdot  sin(\alpha(l))$ becomes ${\tilde c}(n_2-l)=const\cdot sin(l\theta)$
when the variations of $[w+(2n+1)^2]$ at $n\in [n_1,n_2]$ are neglected but it 
is not the exact solution because 
\begin{eqnarray}
&&{\cal E}{\tilde c}(n_2-l)=\lambda {\tilde c}(n_2-l) -\frac{1}{2}const\cdot cos((\alpha(l))\{[w+(2(n_2-l)+3)^2-\lambda^2]^{1/2}-\nonumber\\
&&[w+(2(n_2-l)+1)^2-\lambda^2]^{1/2}\}
\label{EB} 
\end{eqnarray} 

\subsection{Classical equations of motion}
\label{motion}

As already noted, Eq. (\ref{estimation}) gives the estimation of the width of the relative dS momentum if the mass of particle 2 is much greater
than the mass of particle 1. It also follows from Eq. (\ref{G}) that not only $p$ is a very large number but even $lnp$ is very large. Suppose
now that $p$ changes. We do not say that $p$ changes with time because time is a classical notion while we are considering a pure quantum problem.
Below we propose a scenario that classical time arises as a consequence of the fact that $p$ changes. As noted in Ref. \cite{monograph}, there are
reasons to think that at early stages of the Universe $p$ was much less than now i.e. $p$ is increasing.

If $p$ changes by $\Delta p$ then $\Delta p$ cannot be infinitely small because, roughly speaking, $p$ is an integer. Moreover, a possible scenario is that
at every step $p$ is multiplied by a number $k$ and if $k\gg 1$ then $\Delta p\gg p$. However, in that case $lnp$ changes by $\Delta lnp=lnk$. This
quantity also cannot be infinitely small but it is possible that $\Delta lnp/lnp$ is a very small real number.
As follows from Eq. (\ref{estimation}), 
$\Delta\delta /\delta = \Delta lnp /lnp$. Therefore $\Delta\delta /\delta$ does not depend on the heavy mass and depends only on the change of $p$. Since time is a dimensionful parameter, we {\it define} time such that its variation is given by $\Delta t=R \Delta lnp /lnp$. In that case $\Delta t$ also cannot be infinitely small but can be very small in
comparison with macroscopic times. 

In view of Eq. (\ref{G}) and the definition of time the following problem arises. If $p$ changes then does it mean that $G$ changes? In our approach the number $p$ is fundamental while $G$ is not. In view of the remarks in Secs. \ref{symmetry} and \ref{S9}, a problem also arises whether dimensionful quantities can be fundamental. In particular,
as noted in Sec. \ref{S9}, the quantity $G_{dS}$ given by Eq. (\ref{GLambda}) is more fundamental than $G$ because it is dimensionless. Equation (\ref{G}) shows
that $G$ depends not only on $p$ but also on $R$. This parameter has the dimension of meter because people want to deal with Poincare momenta
and not with dimensionless dS angular momenta. So it is not even clear whether $R$ expressed in meters 
changes or not. In any case, among the constants which
are treated as fundamental, $G$ is measured with the least accuracy and its value is known only for approximately 300 years. 
If $\Delta lnp\ll lnp$ then it is quite possible that the change of $G$ could not be noticed for such a short period of time. 
In view of these remarks we assume that relative variations of such quantities as $R$ and $\delta$ are much smaller than
relative variations of standard momenta and coordinates characterizing the particle under consideration. In what follows we use $p$ to denote the magnitude of standard momentum.

The problem arises how $n_2$ changes with the change of $\delta$. Understanding this problem is very difficult because, 
as discussed
in the preceding subsection, even understanding the behavior of the semiclassical WF is very difficult. 
For this reason we can only make
assumptions about the dependence of the variation of $n_2$ on the variation of $\delta$. Since the choice of
the WF is defined by the choice of $c(n_2-1)$ and $c(n_2-1)$ is a function of $\lambda$, we assume that 
$\lambda$ is the conserved quantity. For simplicity, in what follows we will write $n$ instead of $n_2$ and consider only nonrelativistic approximation.

Consider a situation in standard theory when a particle is moving along the $z$-axis
and is attracted or repulsed by a body in the origin. Consider first a possibility that
\begin{equation}
\Delta n=\pm (W-\lambda^2)^{1/2}\frac{\Delta\delta}{2\delta}
\label{1st}
\end{equation}
where the sign depends on whether the particle momentum and radius-vector are parallel or anti-parallel.
We treat Eq. (\ref{1st}) as an approximate consequence of FQT formulated in terms of real numbers and so we can use classical mathematics for
treating this expression with a good approximation.

If $\theta$ is {\it defined} such that $cos\theta=\lambda /W^{1/2}$ and $sin\theta$ is positive
then $\theta\approx sin\theta=(1-\lambda^2/W)^{1/2}$ and
\begin{equation}
\lambda \approx \pm W^{1/2}(1-\theta^2/2)\approx 2R(m+p^2/2m-m\theta^2/2),\quad p\Delta p=m^2 \theta\Delta \theta
\label{1stB}
\end{equation}
The last relation follows from the fact that $\lambda$ is a conserved quantity. Finally, if we {\it define} $r=R\theta$ 
and  note that $n=Rp$
then, as follows from the definition of time and Eqs. (\ref{1st}) and (\ref{1stB})
\begin{equation}
\Delta p=\pm \frac{mr}{R^2}\Delta t,\quad \Delta r=\pm \frac{p}{m}\Delta t
\label{1stC}
\end{equation}
In view of the remarks on Eq. (\ref{coordmom}), the second expression shows that the quantity $r$ defined above
indeed has the meaning of the coordinate. Since the quantities $p$ and $r$ are positive by construction, it is
clear that in our one-dimensional model the sign is $\pm$ when the momentum and radius-vector are collinear
and anticollinear, respectively.

In the approximation when $\Delta t$ in Eq. (\ref{1stC})  can be treated as infinitely small, we get 
$\dot{p}=\pm mr/R^2$, $\dot{r}=\pm p/m$, 
i.e. exactly the Hamilton equations obtained from the Hamiltonian $H=p^2/(2m) - mr^2/(2R^2)$. It
follows from these relations that $\ddot{r}=r/R^2$ in agreement with Eq. (\ref{accel}) (taking into account
that we work in units where $c=1$). Therefore we have repulsion as it should be in accordance with the
consideration in Sec. \ref{2particles}. Here it has been noted that the result for dS antigravity is
compatible with the prescription of standard quantum theory that the coordinate and momentum representations
should be related to each other by the Fourier transform.

Consider now a possibility that
\begin{equation}
\Delta n=\pm \frac{(W-\lambda^2)^2}{4const^2W^{3/2}}\Delta \delta
\label{grav}
\end{equation}
where $const$ is the same as in Eq. (\ref{deltaj}). We can define $\theta$, assume 
that $\theta\ll 1$  and use Eq. (\ref{1stB}) as above. 
Then $\Delta n=\pm W^{1/2}\theta^4\Delta \delta /(4const^2)$. However, if we define $r$ as above then this
quantity will not satisfy the second condition in Eq. (\ref{1stC}), i.e. it will not have the meaning of
coordinate. Therefore in the given case the momenta and coordinates cannot be related by the Fourier transform.
In accordance with Sec. \ref{2particlesdiscrete}, we now {\it define} $\theta=const /(\delta \varphi)^{1/2}$
where $\varphi=r/R$. Then as follows from the definition of time and Eqs. (\ref{deltaj}) and (\ref{1stB})
\begin{equation}
\Delta p=\pm \frac{MmG}{r^2}\Delta t,\quad \Delta r=\mp \frac{p}{m}\Delta t
\label{grav2}
\end{equation}
where $M$ is the mass of the heavy particle 2. As follows from the second expression, the quantity $r$ has now the
meaning of the coordinate in view of the remarks on Eq. (\ref{coordmom}). We conclude that the sign in 
Eq. (\ref{grav}) should be opposite to that in Eq. (\ref{1st}): it is $\pm$ when the momentum and radius-vector are 
anticollinear and collinear, respectively. 
In the approximation when $\Delta t$ is infinitely small we get $\dot{p}=\pm MmG/r^2$, $\dot{r}=\mp p/m$ and
$\ddot{r}=-MG/r^2$. The last relation shows that in this case we have attraction as it should be for gravity.

We have considered two cases when $\Delta n$ is given by Eqs. (\ref{1st}) and (\ref{grav}), respectively. The first case
reproduces standard dS antigravity and the second case --- standard gravity. The comparison of those expressions
shows that the first case takes place when $\delta\theta^3\ll 1$ and the second case --- in the opposite situation
when $\delta\theta^3\gg 1$. As follows from Eq. (\ref{deltaj}), $\delta$ is of the order $R/r_g$ where $r_g$ is the
gravitational radius of the heavy particle 2. As shown above, $\theta=r/R$ in the first case and 
$\theta=const (R/\delta r)^{1/2}\approx (r_g/r)^{1/2}$ in the second one. Therefore the above conditions are
indeed satisfied if $R$ is very large. 

Finally for illustration we consider a possibility to find the solution of the problem of time with the choice of the WF given by
Eq. (\ref{approxA}). Then if ${\tilde c}(n_2,\delta)={\tilde c}(n_2-\delta)$ and 
\begin{equation}
\alpha(n_2,\delta)=\sum_{m=1}^\delta arccos(\frac{\lambda}{[w+(2(n_2-m)+3)^2]^{1/2}})
\label{alpha}
\end{equation}
we have that 
\begin{eqnarray}
&&{\tilde c}(n_2,\delta)=const_1\cdot  cos[\alpha(n_2,\delta)], \nonumber\\
&&{\tilde c}(n_2+\Delta n,\delta+\Delta\delta)=const_2\cdot  cos[\alpha(n_2+\Delta n,\delta+\Delta\delta)]
\label{approxC}
\end{eqnarray} 

As noted in the preceding subsection, it is desirable that the solution ${\tilde c}(n)$ satisfies the condition
${\tilde c}(n_1-1)=0$. For this reason we assume that 
$$cos[\alpha(n_2,\delta)]=cos[\alpha(n_2+\Delta n,\delta+\Delta\delta)]=0.$$
This does not necessarily imply that $\alpha(n_2,\delta)=\alpha(n_2+\Delta n,\delta+\Delta\delta)$ but we
assume that for rather small values of $\Delta n$ and $\Delta\delta$ this is the case. Then
\begin{eqnarray}
&&\sum_{m=\delta+2}^{\delta+\Delta\delta+1} arccos(\frac{\lambda}{[w+(2(n_2+\Delta n-m)+3)^2]^{1/2}})=\nonumber\\
&&-\sum_{m=1}^{\delta+1} [arccos(\frac{\lambda}{[w+(2(n_2+\Delta n-m)+3)^2]^{1/2} })-\nonumber\\
&&arccos(\frac{\lambda}{[w+(2(n_2-m)+3)^2]^{1/2}})]
\label{Delta}
\end{eqnarray}
If $\Delta n\ll n_2$ then the l.h.s. approximately equals $\delta\theta$ and in the first order correction in $\Delta n$
we have the approximate relation
\begin{equation}
\Delta n=- (W-\lambda^2)\frac{\Delta\delta}{4\delta n_2}
\label{Delta2}
\end{equation} 
The r.h.s of this relation differs from the r.h.s. of Eq. (\ref{1st}) by the factor $(W-\lambda^2)^{1/2}/(2n_2)$.
This factor can be greater or less than unity but the solution (\ref{Delta2}) is unacceptable because in this case 
it is not possible to define $r$ satisfying Eq. (\ref{coordmom}). Nevertheless we believe that this example gives
hope that our conjecture on the problem of time can be substantiated with the exact solution of Eq. (\ref{1E}).

\section{Conclusion}
\label{conclusion}

As noted in Refs. \cite{monograph,lev2} and Sec. \ref{fundamentaltheories}, from
the point of view of quantum theory, the notions of infinitely small/large, continuity etc. are unnatural.
As explained in Sec. \ref{mainstatement}, {\it even in standard quantum theory all quantum states can be described with 
any desired accuracy by using only integers}. As noted in Sec. \ref{mainstatement}, from the point of view of describing
quantum states, Hilbert spaces in standard quantum theory contain a big redundancy of elements, 
and rational and real numbers play only the auxiliary role.  

Moreover, as proved in Sec. \ref{mainstatement}, continuous mathematics itself is a special degenerate case of finite mathematics: 
the latter becomes the former in the formal limit when the characteristic $p$ of the ring or field in finite
mathematics goes to infinity. Continuous mathematics describes many data with high accuracy as a consequence of the fact that at the present 
stage of the Universe the characteristic is very large. There is no doubt that the technique of continuous mathematics is useful in many practical 
calculations with high accuracy. However, from the above facts it is clear that the problem of substantiation of this mathematics (which was discussed 
by many famous mathematicians, which has not been solved so far and which probably cannot be solved (e.g. in view of G\"{o}del's incompleteness theorems)) 
is not fundamental because continuous mathematics itself, being a special degenerate case of finite mathematics, is not fundamental.

A natural generalization of standard quantum theory is such that quantum states are elements of a space over
a finite ring or field with the characteristic $p$, and no physical quantity exceeds $p$. In that case the description of 
states does not contain redundant elements, i.e. this description is much more thrifty than in standard quantum theory.
In my discussions with physicists
some of them said that such a generalization is not fundamental because $p$ is simply a cutoff. This point of view
is not correct because finite mathematics cardinally differs 
from standard one and, as follows from the results of Sec. \ref{mainstatement}, fundamental
quantum theory should involve a finite quantity $p$. A historical analogy is that Special Relativity cannot be treated 
simply as classical mechanics with the cutoff $c$ for velocities. 

In Secs. \ref{intro} we argue that, although quantum theory exists for more than 90 years
and it is now clear that classical physical intuition typically does not work here, quantum theory inherited many
its notions from classical theory. Quantum theory is treated as more general as classical one and at some
conditions quantum theory should reproduce all results of classical theory, including classical equations of
motions. However, in quantum theory the notion of space-time is unnatural and should not be
present at all. In addition, quantum theory is based on the results of classical mathematics
developed mainly when people did not know about the existence of atoms and elementary particles. 
It is also known that classical mathematics has foundational problems by its own nature (as follows, for example, from G\"{o}del's incompleteness theorems). 

Let us note that in physics the quantities ($c,\hbar,G$) are usually called fundamental constants but there is no proof that those quantities are always the same in the history of the Universe The quantity $c$ is special because in the present system of units the meter and second
are not independent but are related to each other by {\it the artificial requirement} that $c$ does not change with time. As noted in Sec. \ref{fundamentaltheories},
relativistic theory itself does not need the dimensionful quantity $c$. 

In Refs. \cite{lev4,lev2} we called the quantity $p$ the constant but in those references the
problem of time has not been discussed. A possibility that $p$ may be related to time has been first discussed in Ref. \cite{monograph}. 
Although the number $p$ is a fundamental
parameter defining physical laws, this does not necessarily mean that this number is always the same in 
the history of Universe. We do not say that the number is the same at all times because time is
a pure classical notion and should not be present in quantum theory. Our conjecture is that {\it the existence of classical time 
is a consequence of the fact that $p$ changes} and in Sec. \ref{classicalFQT} we {\it define} time such that its 
variation $\Delta t$ is related to the variation of $p$ as
\begin{equation}
\Delta t=\frac{R}{c}\frac{\Delta lnp}{lnp}
\label{timedef}
\end{equation}
where $R$ is the de Sitter (dS) radius. Then as shown in Subsec.
\ref{motion}, there exist scenarios when classical equations of motions for cosmological acceleration and gravity 
can be obtained from pure quantum
notions without using space, time and standard semiclassical approximation.

In this scenario the goal of quantum theory is to determine how mean values of dS angular momenta change when 
the widths of their distributions change. As shown in Subsec. \ref{1dim}, even in the one-dimensional model
discussed in this subsection the problem of finding exact solutions is very difficult. However, in Subsec. \ref{motion}
we consider two possibilities when classical equations of motion in standard dS antigravity and standard
gravity can be indeed obtained from pure quantum theory without involving any classical notions and standard semiclassical
approximation. 
   
 {\it Acknowledgement:} Volodya Nechitailo told me the idea that the number $p$ may be related to time. I am also grateful to Bernard Bakker, Jos{\'e} Manuel Rodriguez Caballero, Dmitry Logachev 
and Harald Niederreiter for important remarks.


\begin{thebibliography}{99}
\bibitem{textbooks} Lidl R. and Niederreiter H., {\it Finite Fields}, London: Addison-Wesley Publishing Company, 1983; Ireland K. and Rosen M., {\it A Classical Introduction to Modern Number Theory}, New York: Springer-Verlag, 1990; 
Davenport H., {\it The Higher Arithmetic}, Cambridge: University Press, 1999; Zhe-Xian Wan, {\it Lectures on Finite Fields and Galois Rings}, New Jersey: Worlds Scientific, 2003.
\bibitem{lev4} Lev, F., Modular Representations as a Possible Basis of Finite Physics,  {\it J. Math. Phys.}, 1989, vol. 30, no. 9,
pp. 1985-1998; Finiteness of Physics and its Possible Consequences, {\it J. Math. Phys.}, 1993, vol. 34, no. 2, pp. 490-527;
Why is Quantum Theory Based on Complex Numbers?, {\it Finite Fields and Their Applications.}, 2006, vol. 12, no. 3, pp. 336-356;  Quantum Theory and Galois Fields, {\it Int. J. Modern Phys.}, 2006, vol. B20, nos. 11, 12, 13, pp. 1761-1777.
\bibitem{monograph} Lev, F., Finite Quantum Theory and Applications to Gravity and Particle Theory, 2016, {\it arxiv:1104.4647}.
\bibitem{lev2} Lev, F., Why Finite Mathematics Is The Most Fundamental And Ultimate Quantum
Theory Will be based on Finite Mathematics, {\it Physics of Particles and
Nuclei Letters}, 2017, vol. 14, no. 1, pp. 77-82.
\bibitem{Dyson} Dyson F. G., Missed Opportunities. {\it  Bull. Amer. Math. Soc.}, 1972, vol. 78 , no. 5, pp. 635--652.
\bibitem{JPA} Lev, F. M., Could Only Fermions Be Elementary?, {\it J. Phys. A: Mathematical and Theoretical,} 2004, vol. 37, no. 9, pp. 3287-3304.
\bibitem{verificationism}  Misak, C. J., {\it Verificationism: Its History and Prospects}, New York: Routledge, 1995;
Ayer, A.J., {\it Language, Truth and Logic}, in {\it Classics of Philosophy},  New York - Oxford: Oxford University Press, 
1998, pp. 1219-1225; 
 William, G., {\it Lycan$'$s  Philosophy of Language: A Contemporary Introduction}, New York: Routledge, 2000. 
\bibitem{Grayling} Grayling, A.C., {\it Ideas That Matter}, New York: Basic Books,  2012.
\bibitem{Popper} Karl Popper, in {\it Stanford Encyclopedia of Philosophy}.
\bibitem{Zassenhaus} Zassenhaus, H., The Representations of Lie Algebras of Prime Characteristic, {\it Proc. Glasgow Math. Assoc.}, 1954, vol. 2, no. 1, pp. 1-36.
\bibitem{Saniga} Planat, M., Giorgetti, A.,  and Saniga, M.,  Quantum contextual finite geometries from dessins d\'enfants.,
{\it International J. of Geometric Methods in Modern Physics}, World Scientific, 2015, vol. 12, no. 7, 1550067 (18 pages);
Vourdas, A., {\it Finite and Profinite Quantum Systems}, Springer Verlag, XIII, 196pp., 2017. 
\bibitem{Bogolubov} Bogolubov, N. N., Logunov, A. A., Oksak, A. I.,  and  Todorov, I.T.,
{\it General Principles of Quantum Field Theory}, Moscow: Nauka, 1987.
\bibitem{Pauli} Pauli, W., {\it General Principles of Quantum Mechanics}. Berlin: Springer-Verlag,  1980.
The original German edition: {\it Handbuch der Physik}, vol. 5, {\it Prinzipien der Quantentheorie}, Berlin: Springer-Verlag, 1958.
\bibitem{Leon} Leon, J.,  and Maccone,  L., The Pauli objection, {\it Foundations of Physics}, 2017, vol. 47, no. 12, pp. 1597-1608.
\bibitem{Anderson} Anderson, E., Problem of Time in Quantum Gravity, {\it  Annalen der Physik}, 2012, vol. 524, no. 12, pp. 757-786. 
\bibitem{RovelliBook} Rovelli, C., {\it Quantum Gravity.}, Cambridge: Cambridge University Press, 2004.
\bibitem{RovelliTime} Rovelli, C., Forget Time, {\it The FQXI Essay Contest "The Nature of Time"}, 2008.
\bibitem{Keaton} Keaton, G., What is Time?, {\it The FQXI Essay Contest "The Nature of Time"}, 2008.
\bibitem{Pauli2} Pauli, W., The Connection Between Spin and Statistics, {\it Phys. Rev.}, 1940, vol. 58, no. 8, pp. 716-722.
\bibitem{NW} Newton, T.D., and Wigner, E.P., Localized States for Elementary Systems, {\it Rev. Mod. Phys.}, 1949, vol. 21, no. 3, pp. 400-405.
\bibitem{Darwin} Darwin, C. G., Free Motion in the Wave Mechanics, {\it Proc. R. Soc. London}, 1927, vol. A117, issue 776, pp. 258-293.
\bibitem{lev1} Lev, F.,  Fundamental Quantal Paradox And Its Resolution, {\it Physics of Particles and
Nuclei Letters}, 2017, vol. 14, no. 3, pp. 444-452.
\bibitem{PRD} Lev, F., de Sitter Symmetry and Quantum Theory, {\it Phys. Rev.}, 2012, vol. D85, 065003.
\bibitem{Dir} Dirac, P., A., M., Forms of Relativistic Dynamics, {\it Rev. Mod. Phys.}, 1949, vol. 21, no. 3, pp. 392-399.
\bibitem{Wigner} Wigner, E.P., On Unitary Representations of the Inhomogeneous Lorentz Group,
{\it Ann. Math.}, 1939, vol. 40, no. 1, pp. 149-204.
\bibitem{dS} Lev, F., The Problem of Interactions in de Sitter Invariant Theories, {\it J. Phys.}, 1999, vol. A32, no. 7, pp. 1225-1239. 
\bibitem{Perlmutter} Perlmutter, S., Aldering, G., Goldhaber, G., {\it et. al.}, Measurement of Omega and 
Lambda from H42 High-redshift Supernovae, {\it Astrophys. J.}, 1999, vol. 517, no. 2, pp. 565-586.
\bibitem{Mensky} Mensky, M., B., {\it Method of Induced Representations.
Space-time and Concept of Particles}, Moscow: Nauka, 1976.
\bibitem{Bianchi} Bianchi E., and Rovelli, C., Why all These Prejudices Against a Constant?, 2010,
{\it arXiv:1002.3966v3 (astro-ph.CO)}.
\end{thebibliography}
\end{document}